\documentclass[fleqn,10pt]{SelfArx} 

\definecolor{color1}{RGB}{0,0,90} 
\definecolor{color2}{RGB}{0,20,20} 

\usepackage{hyperref} 
\hypersetup{hidelinks,colorlinks,breaklinks=true,urlcolor=color2,citecolor=color1,linkcolor=color1,bookmarksopen=false,pdftitle={Title},pdfauthor={Author},urlcolor=blue}

\usepackage{graphicx}
\usepackage[square]{natbib}
\usepackage{amsmath}
\usepackage{multirow}
\usepackage{subcaption}
\usepackage{pdflscape}
\usepackage[final]{pdfpages}

\usepackage{graphicx}
\usepackage{natbib}
\usepackage{amsmath}
\usepackage{amsfonts}
\usepackage{multirow}
\usepackage{tikz}
\usetikzlibrary{calc}

\DeclareMathOperator{\sgn}{sgn}
\DeclareMathOperator{\atanh}{atanh}
\DeclareMathOperator{\atan}{atan}

\usepackage{mathtools,amssymb,lipsum}

\usepackage{cuted}
\usepackage{boondox-calo}

\usepackage{calligra}

\DeclareMathAlphabet{\mathcalligra}{T1}{calligra}{m}{n} \DeclareFontShape{T1}{calligra}{m}{n}{<->s*[2.2]callig15}{}

\newcommand{\n}[1]{\mathrm{#1}}

\newcommand{\ud}{\mathrm{d}}

\newcommand{\zhat}{\mathbf{\hat{z}}}
\newcommand{\nhat}{\mathbf{\hat{n}}}

\newcommand{\bfr}{\mathbf{r}}
\newcommand{\bfM}{\mathbf{M}}
\newcommand{\bfH}{\mathbf{H}}
\newcommand{\bfB}{\mathbf{B}}
\newcommand{\bfN}{\mathbb{N}}

\newcommand{\uhat}{\hat{\mathbf{u}}}
\newcommand{\bfD}{\mathbb{D}}
\newcommand{\Rb}{R_\n{b}}
\newcommand{\cA}{\mathcal{A}}
\newcommand{\cB}{\mathcal{B}}
\newcommand{\cC}{\mathcal{C}}
\newcommand{\cD}{\mathcal{D}}
\newcommand{\cL}{\mathcal{L}}
\newcommand{\cP}{\mathcal{P}}
\newcommand{\cQ}{\mathcal{Q}}
\newcommand{\cR}{\mathcal{R}}
\newcommand{\cU}{\mathcal{U}}
\newcommand{\cV}{\mathcal{V}}
\newcommand{\cW}{\mathcal{W}}

\JournalInfo{~} 
\Archive{~} 

\PaperTitle{The magnetic field of a homogeneously magnetized cone} 

\Authors{Rasmus Bjørk} 
\affiliation{\textit{Department of Energy Conversion and Storage, Technical University of Denmark - DTU, DK-2800 Kgs. Lyngby, Denmark}} 
\affiliation{*\textbf{Corresponding author}: rabj@dtu.dk} 

\newcommand{\keywordname}{Keywords} 

\Abstract{The magnetic field of a homogeneously magnetized right circular cone is determined. The field is expressed as the product between the demagnetization tensor describing the geometrical part of the problem and the magnetization. Additionally, its magnetic scalar potential is expressed as the product between a demagnetization vector and the magnetization. All nine components of the tensor and three components of the vector are expressed in terms of elliptic integrals of the first, second and third kind, and a single scalar function of the field point. This function is given as a one-dimensional integral and as a series, and it reduces to the known logarithm on the axis of the cone. The demagnetization tensor and vector for the truncated cone and the annular cone are given by simple superposition. The magnetometric demagnetization tensor is also derived in closed form. The tensors are verified against direct numerical integration of the surface charges and with a finite element model, showing excellent agreement.}

\begin{document}

\flushbottom 

\maketitle 


\thispagestyle{empty} 

\section{Introduction}
\label{sec:intro}
Magnetic fields are used in a large variety of applications such as permanent magnet motors and generators \citep{Zhu_1993,Markovic_2009}, magnetic gears and bearings \citep{Lubin_2010,Ravaud_2009a} and nuclear magnetic resonance (NMR) systems \citep{Moresi_2003,Raich_2004}. The magnetic field generated by a permanent magnet or a system of these is often calculated using a finite element method (FEM) approach, which is limited in precision and speed by the fineness of the finite element mesh. However, if the magnet is uniformly magnetized, the magnetic field can be calculated through the demagnetization tensor, $\bfN$, which expresses the relation between the magnetic field and the magnetization, as
\begin{equation}
\bfH(\mathbf{r})=-\bfN\cdot\bfM,
\label{eq:demag}
\end{equation}
where $\bfH$ is the magnetic field at a given point $\mathbf{r}=(x,y,z)$ and $\bfM$ is the magnetization. The demagnetization tensor $\bfN$ is a rank-2 symmetric tensor \citep{Moskowitz_1966}, which is known analytically for a range of geometries, e.g.\ for an ellipsoid \citep{Osborn_1945,Tejedor_1995}, a rectangular prism \citep{Joseph_1965,Smith_2010}, a finite cylinder \citep{Kraus_1973,Caciagli_2018}, a cylindrical tile \citep{Nielsen_2020,Slanovc_2022}, a circular piece \citep{Nielsen_2026}, a tetrahedron \citep{Nielsen_2019}, cylindrical shells \citep{Beleggia_2009} and a hollow sphere \citep{PratCamps_2016}. The demagnetization tensor can also be expressed in Fourier space \citep{Beleggia_2003}. Analogously, the magnetic scalar potential, $\phi_M$, the gradient of which is the magnetic field, $\bfH=-\nabla{}\phi_M$, can be written as the product of a demagnetization vector, $\pmb{\Psi}(\mathbf{r})$, and the magnetization as $\phi_M(\mathbf{r})=\pmb{\Psi}(\mathbf{r})\cdot\mathbf{M}$, which has been done for the rectangular prism \citep{James_2026} and the cylindrical tile \citep{Bjork_2025}.

In this work we consider the magnetic field generated by a homogeneously magnetized circular cone, shown in Fig.~\ref{fig:geometry}. Conically shaped magnets concentrate the magnetic flux towards the apex and are used in nuclear magnetic resonance (NMR) systems, magnetic resonance imaging (MRI) systems and in magnetic field sensors \citep{Nguyen_2020a,Hart_2020}. The field of a cone with an axial magnetization was derived in closed form by Nguyen \citep{Nguyen_2020a}, and was also studied with an arbitrary uniform magnetization also by Nguyen \citep{Nguyen_2020b}, but here no closed form solution was determined. Instead the field is given as integrals over the azimuth of the source point that have to be evaluated numerically. Only for field evaluations on the axis were closed-form expressions given. Hart et al.~\citep{Hart_2020} expanded the Coulomb kernel in toroidal harmonics and obtained the potential and the field of an axially magnetized truncated cone as infinite series of Gauss and Appell hypergeometric functions, again with closed forms on the axis only. Smith et al.~\citep{Smith_2021} treated the lateral surface of the truncated cone as a conical sheet of uniform azimuthal current, which is the equivalent surface current of the axial magnetization, and obtained the off-axis field as integrals of toroidal functions or as series of Appell hypergeometric functions. Batishchev and Hyde \citep{Batishchev_2023} derived the on-axis field of sharp and truncated cones in closed form and used it to optimize the field and its gradient at the apex. The same approach as for the cone, with the azimuthal integral left for numerical evaluation, has been applied to a magnetized paraboloid \citep{Nguyen_2025}, and ring magnets with a trapezoidal cross section, which are bounded by conical surfaces, have been treated numerically by means of magnetic charges \citep{Ivkovic_2025}. The magnetometric demagnetization factors of cones, bicones and cones truncated by a spherical cap have been obtained numerically or as series expressions \citep{Tandon_2004,Phatak_2011,Prozorov_2018}.

In this work we derive the analytical expressions in closed form for the demagnetization tensor and for the demagnetization vector of a homogeneously magnetized cone at any point inside or outside the cone. The derivation is carried out for a transverse magnetization, for which only the lateral surface of the cone is charged, as the symmetry and the trace of the tensor then supply the remaining column, which is the field of the axially magnetized cone. We show that all components reduce to elementary functions and complete elliptic integrals of the first, second and third kind, with the exception of a single scalar function, the azimuthal average of the potential of a generator of the cone, i.e.\ of a straight line on the lateral surface from the apex to the rim of the base. We show that this function cannot be reduced to elliptic integrals and give it as a one-dimensional integral and as a series. We also derive the magnetometric demagnetization tensor of the cone in closed form, which depends on the cone half-opening angle alone. The derived expressions are validated against numerical integration as well as against comparison against a finite element model. The derived expressions are provided as Matlab and python scripts as part of the MagTense open source simulation framework \cite{MagTense,Bjork_2021}.

\section{The magnetic field of a cone}
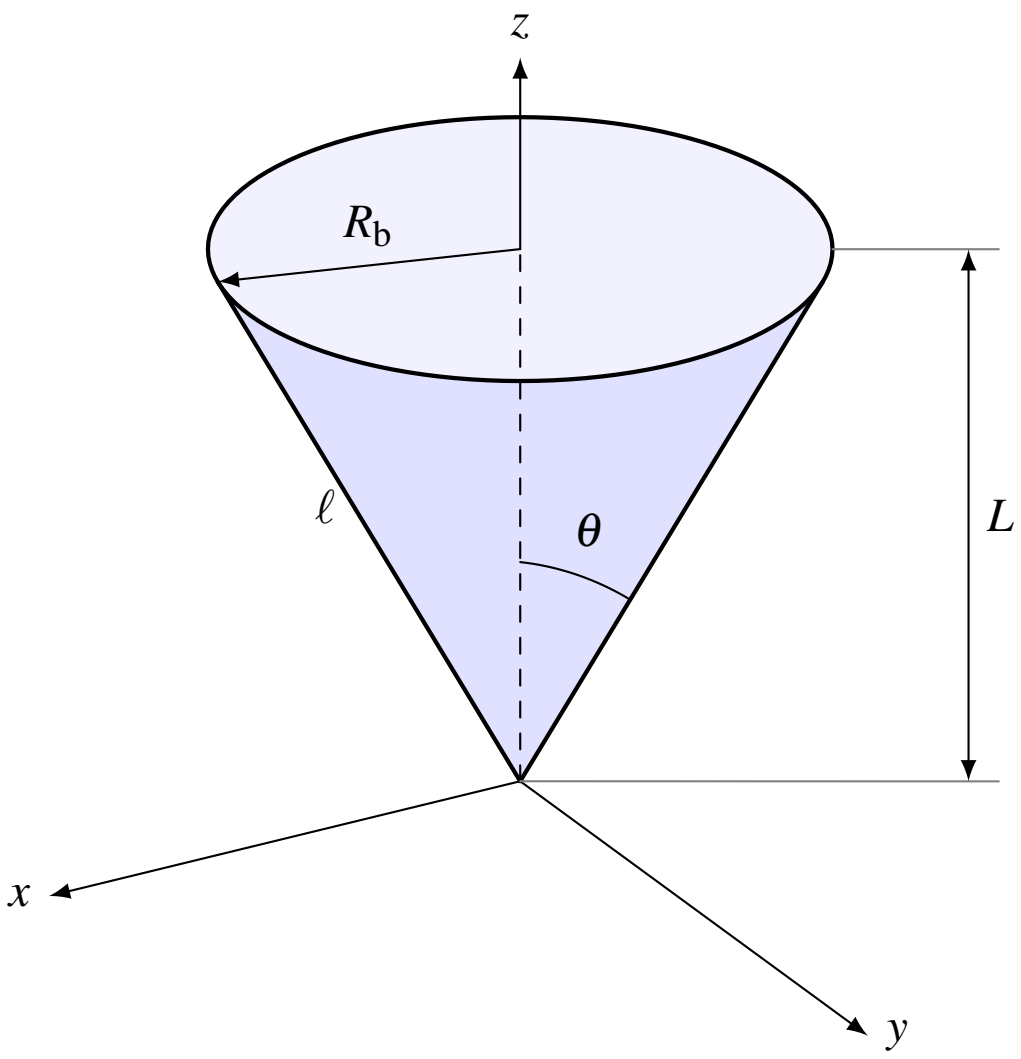
\begin{figure}[!ht]
\centering
\begin{tikzpicture}[>=latex,font=\small,line cap=round,line join=round,
  x={(-0.866cm,-0.211cm)},y={(0.500cm,-0.366cm)},z={(0cm,0.906cm)}]
\def\conetheta{28}                                   
\def\coneell{4.4}                                    
\pgfmathsetmacro{\coneL}{\coneell*cos(\conetheta)}   
\pgfmathsetmacro{\coneR}{\coneell*sin(\conetheta)}   
\pgfmathsetmacro{\silA}{60+acos(tan(\conetheta)*tan(25))}   
\pgfmathsetmacro{\silB}{60-acos(tan(\conetheta)*tan(25))}   
\draw[->] (0,0,0) -- (3.6,0,0) node[left]{$x$};
\draw[->] (0,0,0) -- (0,4.6,0) node[right]{$y$};
\fill[blue!12] (0,0,0) -- plot[domain=\silB:\silA,samples=60,variable=\t] ({\coneR*cos(\t)},{\coneR*sin(\t)},{\coneL}) -- cycle;
\filldraw[thick,fill=blue!5] plot[domain=0:360,samples=121,variable=\t] ({\coneR*cos(\t)},{\coneR*sin(\t)},{\coneL}) -- cycle;
\draw[thick] ({\coneR*cos(\silA)},{\coneR*sin(\silA)},{\coneL}) -- (0,0,0) -- ({\coneR*cos(\silB)},{\coneR*sin(\silB)},{\coneL});
\draw[dashed] (0,0,0) -- (0,0,\coneL);
\draw[->] (0,0,\coneL) -- (0,0,{\coneL+1.4}) node[above]{$z$};
\draw[->] (0,0,\coneL) -- ({\coneR*cos(\silB)},{\coneR*sin(\silB)},\coneL) node[midway,above]{$\Rb$};
\draw[gray] ({-0.866*\coneR},{0.5*\coneR},\coneL) -- ({-0.866*(\coneR+1.1)},{0.5*(\coneR+1.1)},\coneL);
\draw[gray] (0,0,0) -- ({-0.866*(\coneR+1.1)},{0.5*(\coneR+1.1)},0);
\draw[<->] ({-0.866*(\coneR+0.9)},{0.5*(\coneR+0.9)},0) -- ({-0.866*(\coneR+0.9)},{0.5*(\coneR+0.9)},\coneL) node[midway,right]{$L$};
\draw plot[domain=0:\conetheta,samples=20,variable=\t] ({1.6*sin(\t)*cos(\silA)},{1.6*sin(\t)*sin(\silA)},{1.6*cos(\t)});
\node at ({1.95*sin(0.5*\conetheta)*cos(\silA)},{1.95*sin(0.5*\conetheta)*sin(\silA)},{1.95*cos(0.5*\conetheta)}) {$\theta$};
\node[left] at ({0.55*\coneR*cos(\silB)},{0.55*\coneR*sin(\silB)},{0.55*\coneL}) {$\ell$};
\end{tikzpicture}
\caption{The solid cone with its apex at the origin, its axis along $z$, height $L$, base radius $\Rb$, half-opening angle $\theta$ and slant length $\ell=\sqrt{L^2+\Rb^2}$.}
\label{fig:geometry}
\end{figure}

\label{sec:model}
We consider a solid right circular cone with its apex at the origin and its axis along the $z$-axis, as shown in Fig.~\ref{fig:geometry}. The cone is completely specified by its height $L$ and its base radius $\Rb$. The expressions derived below are most compact in terms of the half-opening angle $\theta$ and the slant length $\ell$ of the lateral surface, which are
\begin{equation}
\begin{aligned}
&\theta=\atan\frac{\Rb}{L},\qquad \ell=\sqrt{L^2+\Rb^2},\\
&\textrm{so that}\qquad L=\ell\cos\theta,\qquad \Rb=\ell\sin\theta .
\end{aligned}
\label{eq:thetaell}
\end{equation}
The straight segments of length $\ell$ from the apex to the rim of the base, which form the lateral surface, are called the generators of the cone. We use cylindrical coordinates $(\rho,\varphi,z)$, where $\rho=\sqrt{x^2+y^2}$ is the distance from the $z$-axis and $\varphi$ is the azimuth measured from the $x$-axis, so that a point is $\bfr=(x,y,z)=(\rho\cos\varphi,\rho\sin\varphi,z)$, see also Fig.~\ref{fig:distances}. The base of the cone is then the disk $\rho\le\Rb$ in the plane $z=L$. The distance of a point from the apex is denoted
\begin{equation}
c=|\bfr|=\sqrt{\rho^2+z^2} .
\end{equation}

We assume the cone to be homogeneously magnetized with the magnetization $\bfM=(M_x,M_y,M_z)$. We wish to determine the magnetic field, $\bfH$, and the magnetic scalar potential, $\phi$, at any point $\bfr=(x,y,z)$ inside or outside the cone. We assume no free currents, which makes the magnetic field irrotational and allows it to be written as the gradient of the magnetic scalar potential as $\bfH=-\nabla\phi_M$. For a homogeneously magnetized body the magnetic charge volume density, $\rho_M=-\nabla\cdot\bfM$, vanishes, and the scalar potential is produced only by the magnetic charge surface density, $\sigma_M=\nhat\cdot\bfM$, on the enclosing surface $S'$, where $\nhat$ is the outward unit normal of the surface. The magnetostatic scalar potential is \citep{Jackson_1999}
\begin{equation}
\phi_M(\bfr)=\frac{1}{4\pi}\oint_{S'}\frac{\nhat(\bfr')\cdot\bfM}{\|\bfr-\bfr'\|}\,\ud S' .
\label{eq:phi}
\end{equation}
We explicitly note that there are two sets of coordinates. The coordinates marked with a $'$ are the coordinates of the surface that creates the magnetic field, whereas the non-marked coordinates are those of the point at which the field is evaluated. As the magnetization is constant it can be moved outside the integral, and the potential can be written as the dot product of a demagnetization vector, which contains all the geometric information, and the magnetization \citep{James_2026,Bjork_2025},
\begin{equation}
\phi_M(\bfr)=\pmb{\Psi}(\bfr)\cdot\bfM,\qquad
\pmb{\Psi}(\bfr)=\frac{1}{4\pi}\oint_{S'}\nhat(\bfr')\,D\,\ud S' ,
\label{eq:Nphi}
\end{equation}
with
\begin{equation}
D(\bfr')=\frac{1}{\|\bfr-\bfr'\|} .
\label{eq:Ddef}
\end{equation}
The magnetic field is then obtained as $\bfH=-\nabla\phi_M$. As the gradient operator and the integration are with respect to different sets of coordinates, the order of these is interchangeable, and we get
\begin{equation}
\begin{aligned}
\bfH(\bfr)&=-\bfN(\bfr)\cdot\bfM,\\
N_{ij}(\bfr)&=\frac{1}{4\pi}\oint_{S'} n'_j\,\partial_i D\,\ud S'=\frac{\partial \Psi_{j}}{\partial r_i} ,
\end{aligned}
\label{eq:Ndef}
\end{equation}
where $\partial_i$ denotes the partial derivative with respect to the $i$-th coordinate of the field point, $i=x,y,z$, and $n'_j$ is the $j$-th component of $\nhat(\bfr')$. The tensor is symmetric and its trace is $1$ for field points inside the body and $0$ outside \citep{Moskowitz_1966}.

The components of the tensor are computed by integrating along the entire closed surface of the cone. There are two surfaces, the lateral surface and the base, and only the column of $\bfN$ that multiplies the magnetization component along the surface normal is non-zero for a given surface. The total tensor is the sum of the contributions from the two surfaces,
\begin{equation}
\bfN=\bfN^\n{Lateral}+\bfN^\n{Base} ,
\label{eq:Nsum}
\end{equation}
and likewise for the demagnetization vector.

\subsection{Surface 1 -- the lateral surface}
A point on the lateral surface is $\bfr'=s\,\uhat(\varphi')$, with $0\le s\le\ell$ being the slant coordinate measured from the apex, and $0\le\varphi'<2\pi$ being the azimuth of the source point and
\begin{eqnarray}
\uhat(\varphi')&=&\left(\begin{array}{c}\sin\theta\cos\varphi'\\ \sin\theta\sin\varphi'\\ \cos\theta\end{array}\right),\;
\nhat(\varphi')=\left(\begin{array}{c}\cos\theta\cos\varphi'\\ \cos\theta\sin\varphi'\\ -\sin\theta\end{array}\right),\nonumber\\
\ud S'&=&s\sin\theta\,\ud s\,\ud\varphi' ,
\label{eq:surface}
\end{eqnarray}
the unit vector along the slant generator and the outward normal. The magnetic charge surface density is $\sigma_M=\cos\theta\,(M_x\cos\varphi'+M_y\sin\varphi')-M_z\sin\theta$, and the elements of the tensor are thus
\begin{align}
N^\n{Lateral}_{ix}&=\frac{\sin\theta\cos\theta}{4\pi}\int_0^{2\pi}\!\!\int_0^{\ell}\cos\varphi'\,\partial_i D\,s\,\ud s\,\ud\varphi',\nonumber\\
N^\n{Lateral}_{iy}&=\frac{\sin\theta\cos\theta}{4\pi}\int_0^{2\pi}\!\!\int_0^{\ell}\sin\varphi'\,\partial_i D\,s\,\ud s\,\ud\varphi',\nonumber\\
N^\n{Lateral}_{iz}&=-\frac{\sin^2\!\theta}{4\pi}\int_0^{2\pi}\!\!\int_0^{\ell}\partial_i D\,s\,\ud s\,\ud\varphi' .
\label{eq:latint}
\end{align}
for $i=x,y,z$.

\subsection{Surface 2 -- the base}
The base is the disk $\rho'\le\Rb$ in the plane $z'=L$ with the normal $\nhat'=\zhat$, so that $\sigma_M=M_z$ and
\begin{equation}
N^\n{Base}_{iz}=\frac{1}{4\pi}\int_0^{2\pi}\!\!\int_0^{\Rb}\partial_i D\,\rho'\,\ud\rho'\,\ud\varphi'\Big|_{z'=L} ,
\label{eq:baseint}
\end{equation}
and $N^\n{Base}_{ix}=N^\n{Base}_{iy}=0$. This integral is identical to that over an end surface of the finite cylinder which has a closed-form analytical solution \citep{Caciagli_2018}.

\section{Evaluating the integrals}
\label{sec:evaluating}
We first evaluate $N^\n{Lateral}_{ix}$ and $N^\n{Lateral}_{iy}$, i.e. the two columns of the complete tensor for a transverse magnetization. Then the last column, $N^\n{Lateral}_{iz}$, follows from the symmetry of the tensor and the trace rule. In Eq.~\eqref{eq:latint} the integration along the slant generator, i.e.\ over $s$, is elementary, and the remaining integration over the azimuth of the source point is the step that has so far been done numerically \citep{Nguyen_2020b}.

\subsection{Integration along the generators}
\label{sec:generators}
\begin{figure*}[!ht]
\centering
\def\conetheta{28}                                   
\def\coneell{4.4}                                    
\def\fieldrho{4.0}\def\fieldphi{20}\def\fieldz{2.6}  
\def\srcphi{75}                                      
\def\srcs{2.4}                                       
\pgfmathsetmacro{\coneL}{\coneell*cos(\conetheta)}
\pgfmathsetmacro{\coneR}{\coneell*sin(\conetheta)}
\pgfmathsetmacro{\fieldb}{\fieldz*cos(\conetheta)+\fieldrho*sin(\conetheta)*cos(\srcphi-\fieldphi)}  
\pgfmathsetmacro{\fieldc}{sqrt(\fieldrho*\fieldrho+\fieldz*\fieldz)}                                
\pgfmathsetmacro{\fieldd}{sqrt(\fieldc*\fieldc-\fieldb*\fieldb)}                                    
%
\begin{tikzpicture}[>=latex,font=\small,line cap=round,line join=round,scale=0.74,
  x={(-0.866cm,-0.211cm)},y={(0.500cm,-0.366cm)},z={(0cm,0.906cm)}]
\pgfmathsetmacro{\silA}{60+acos(tan(\conetheta)*tan(25))}
\pgfmathsetmacro{\silB}{60-acos(tan(\conetheta)*tan(25))}
\coordinate (O) at (0,0,0);
\coordinate (P) at ({\fieldrho*cos(\fieldphi)},{\fieldrho*sin(\fieldphi)},\fieldz);          
\coordinate (Pxy) at ({\fieldrho*cos(\fieldphi)},{\fieldrho*sin(\fieldphi)},0);               
\coordinate (S) at ({\srcs*sin(\conetheta)*cos(\srcphi)},{\srcs*sin(\conetheta)*sin(\srcphi)},{\srcs*cos(\conetheta)});  
\coordinate (Q) at ({\coneR*cos(\srcphi)},{\coneR*sin(\srcphi)},\coneL);                      
\draw[->] (O) -- (4.6,0,0) node[left]{$x$};
\draw[->] (O) -- (0,3.6,0) node[right]{$y$};
\fill[blue!12] (O) -- plot[domain=\silB:\silA,samples=60,variable=\t] ({\coneR*cos(\t)},{\coneR*sin(\t)},{\coneL}) -- cycle;
\filldraw[fill=blue!5] plot[domain=0:360,samples=121,variable=\t] ({\coneR*cos(\t)},{\coneR*sin(\t)},{\coneL}) -- cycle;
\draw ({\coneR*cos(\silA)},{\coneR*sin(\silA)},{\coneL}) -- (O) -- ({\coneR*cos(\silB)},{\coneR*sin(\silB)},{\coneL});
\draw[dashed] (O) -- (0,0,\coneL);
\draw[->] (0,0,\coneL) -- (0,0,{\coneL+1.2}) node[above]{$z$};
\draw[densely dashed,gray] (O) -- (Pxy) node[pos=0.6,below right=-2pt,black]{$\rho$};
\draw[densely dashed,gray] (Pxy) -- (P) node[pos=0.4,left,black]{$z$};
\draw[very thick,red!75!black] (O) -- (Q);
\node[right=1pt,inner sep=0.5pt] at ($(O)!0.68!(S)$) {$s$};
\draw (O) -- (P) node[pos=0.5,below=0pt]{$c$};
\draw (P) -- (S) node[pos=0.42,below=0pt]{$\Delta_s$};
\draw (P) -- (Q) node[pos=0.33,above=1pt]{$\Delta_\ell$};
\fill (P) circle (1.6pt) node[left=2pt]{$\bfr$};
\fill[red!75!black] (S) circle (1.6pt);
\node[above right=-1pt] at (S) {$\bfr'$};
\node at (2.0,2.0,-2.3) {(a)};
\end{tikzpicture}
\hfill
\begin{tikzpicture}[>=latex,font=\small,line cap=round,line join=round,scale=0.56]
\filldraw[fill=blue!5] (0,0) circle (\coneR);
\draw[->] (-2.6,0) -- (4.7,0) node[right]{$x$};
\draw[->] (0,-2.6) -- (0,4.2) node[above]{$y$};
\draw[densely dashed,gray] (0,0) -- (\fieldphi:\fieldrho) node[pos=0.94,below right=-3pt,black]{$\rho$};
\draw[densely dashed,gray] (\srcphi:\coneR) -- (\srcphi:3.9);
\draw[very thick,red!75!black] (0,0) -- (\srcphi:\coneR);
\fill (\fieldphi:\fieldrho) circle (2.1pt) node[above right=0pt]{$\bfr$};
\fill[red!75!black] (\srcphi:{\srcs*sin(\conetheta)}) circle (2.1pt) node[left=2pt,black]{$\bfr'$};
\draw[->] (2.45,0) arc (0:\fieldphi:2.45);
\node at ({0.5*\fieldphi}:2.85) {$\varphi$};
\draw[->] (1.0,0) arc (0:\srcphi:1.0);
\node at (40:1.42) {$\varphi'$};
\draw[->] (\fieldphi:3.2) arc (\fieldphi:\srcphi:3.2);
\node at ({0.5*(\fieldphi+\srcphi)}:3.62) {$\psi$};
\draw[->] (0,0) -- (215:\coneR) node[pos=0.55,below right=-2pt]{$\Rb$};
\node at (1.0,-3.9) {(b)};
\end{tikzpicture}
\hfill
\begin{tikzpicture}[>=latex,font=\small,line cap=round,line join=round,scale=0.80]
\coordinate (O) at (0,0);
\coordinate (S) at (\srcs,0);
\coordinate (F) at (\fieldb,0);
\coordinate (Q) at (\coneell,0);
\coordinate (P) at (\fieldb,\fieldd);
\draw[very thick,red!75!black] (O) -- (Q);
\draw[densely dashed,gray] (Q) -- (5.0,0);
\draw[->,thick] (O) -- (0.95,0);
\node[above] at (0.55,0) {$\uhat$};
\draw (O) -- (P) node[midway,above left=-1pt]{$c$};
\draw (S) -- (P) node[pos=0.45,left=0pt]{$\Delta_s$};
\draw (Q) -- (P) node[midway,right=1pt]{$\Delta_\ell$};
\draw[densely dashed] (F) -- (P) node[pos=0.3,right=-1pt]{$d$};
\draw (F) ++(0.22,0) -- ++(0,0.22) -- ++(-0.22,0);
\fill (P) circle (1.6pt) node[above=2pt]{$\bfr$};
\fill[red!75!black] (S) circle (1.6pt) node[above left=0pt,black]{$\bfr'$};
\fill (O) circle (1.2pt);
\fill (Q) circle (1.2pt);
\foreach \xx/\yy/\lab in {\srcs/-0.5/s,\fieldb/-1.05/b,\coneell/-1.6/\ell}{
  \draw[gray] (\xx,0) -- (\xx,{\yy-0.12});
  \draw[<->] (0,\yy) -- (\xx,\yy) node[midway,fill=white,inner sep=1.5pt]{$\lab$};}
\draw[gray] (0,0) -- (0,-1.72);
\node[below] at (0,-1.75) {apex};
\node[below] at (\coneell,-1.75) {rim};
\node at (2.4,-2.85) {(c)};
\end{tikzpicture}
\caption{The quantities used in the integration along the slant generators. (a) The field point $\bfr$ with the cylindrical coordinates $\rho$ and $z$, a slant generator of the cone, shown in red, and a source point $\bfr'$ on it at the slant coordinate $s$. (b) View along the $z$-axis with the azimuth $\varphi$ of the field point, the azimuth $\varphi'$ of the generator and their difference $\psi=\varphi'-\varphi$. (c) The plane through the generator and the field point, with the unit vector $\uhat$ along the generator, the distance $c$ from the apex to the field point, its projection $b$ on the generator, the distance $d$ to the generator, the distance $\Delta_s$ to the source point and the distance $\Delta_\ell$ to the point of the rim.}
\label{fig:distances}
\end{figure*}
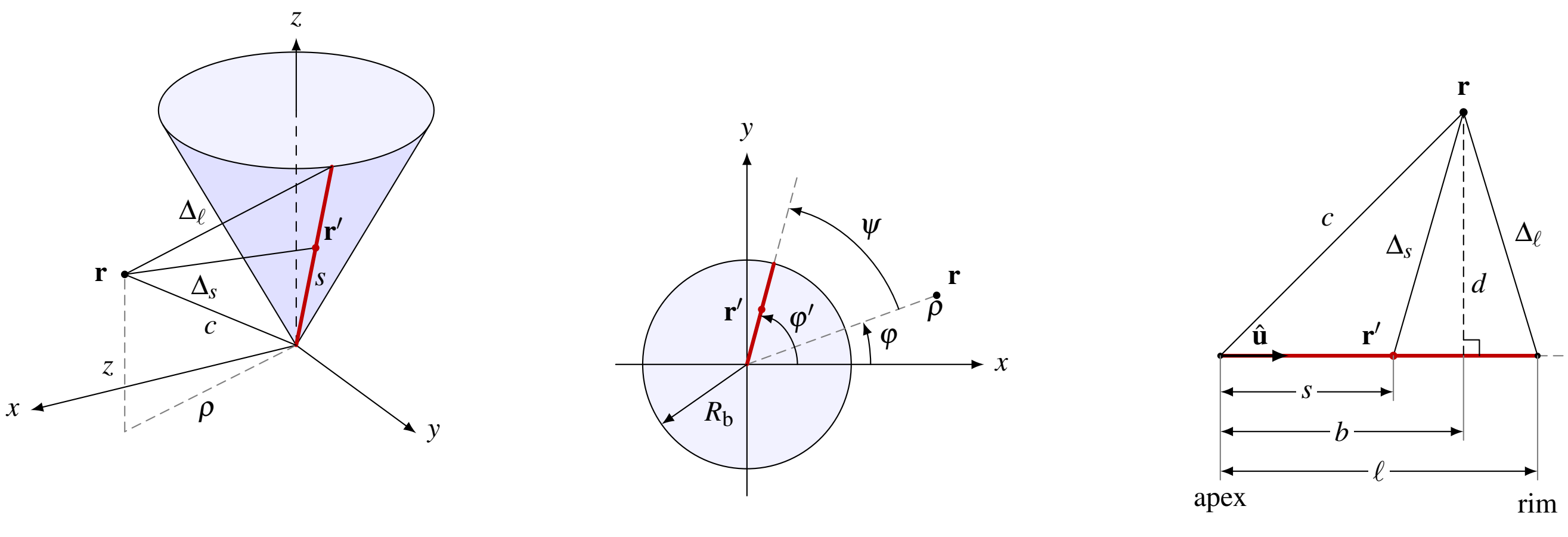

Following Refs.~\citep{Nielsen_2020,Nielsen_2026} the coordinate system is rotated about the $z$-axis by the azimuth $\varphi$ of the field point, so that the field point lies in the $xz$-plane, which replaces the azimuth of the source point by
\begin{equation}
\psi=\varphi'-\varphi .
\end{equation}
The squared distance from the field point to the source point is then $s^2-2bs+c^2$, with
\begin{equation}
b=\bfr\cdot\uhat=B+A\cos\psi,\quad A=\rho\sin\theta,\quad B=z\cos\theta ,
\label{eq:b}
\end{equation}
the projection of the field point on the generator, and $d^2=c^2-b^2$ is the squared distance from the field point to the line that contains the generator. We write $\Delta_s=\sqrt{s^2-2bs+c^2}$ for the distance from the field point to the source point at the slant coordinate $s$, and $\Delta_\ell$ for its value at $s=\ell$, which is the distance to the point of the rim of the base at the azimuth $\varphi'$,
\begin{equation}
\Delta_\ell=\sqrt{\rho^2+\Rb^2-2\rho\Rb\cos\psi+(z-L)^2} .
\label{eq:Delta}
\end{equation}
These angles and distances are shown in Fig.~\ref{fig:distances}.

Inserting $\partial_iD=-(r_i-r'_i)D^3$ in Eq.~\eqref{eq:latint}, the integrals over $s$ are
\begin{align}
\cL&=\int_0^\ell\frac{\ud s}{\Delta_s}=\ln\frac{\ell-b+\Delta_\ell}{c-b}=\ln\frac{c+\Delta_\ell+\ell}{c+\Delta_\ell-\ell},\nonumber\\
J_0&=\int_0^\ell\frac{\ud s}{\Delta_s^3}=\frac{\ell-b}{d^2\Delta_\ell}+\frac{b}{d^2c},\nonumber\\
J_1&=\int_0^\ell\frac{s\,\ud s}{\Delta_s^3}=\frac{c\Delta_\ell+b\ell-c^2}{d^2\Delta_\ell},\nonumber\\
J_2&=\int_0^\ell\frac{s^2\,\ud s}{\Delta_s^3}=\cL+2bJ_1-c^2J_0.
\label{eq:slant}
\end{align}
In terms of these, Eq.~\eqref{eq:latint} becomes
\begin{equation}
N^\n{Lateral}_{ix}=-\frac{\sin\theta\cos\theta}{4\pi}\int_0^{2\pi}\cos\varphi'\left[r_iJ_1-\hat{u}_iJ_2\right]\ud\varphi' ,
\label{eq:Nazimuth}
\end{equation}
and correspondingly with $\sin\varphi'$ for the second column. This is the form given in Ref.~\citep{Nguyen_2020b}.

\subsection{The azimuthal integrals}
\label{sec:azimuth}
The functions $J_0$, $J_1$ and $\cL$ depend on $\psi$ only through $\cos\psi$ and are therefore even in $\psi$, while $\cos\varphi'=\cos\psi\cos\varphi-\sin\psi\sin\varphi$ and $\sin\varphi'=\sin\psi\cos\varphi+\cos\psi\sin\varphi$. Terms odd in $\psi$ integrate to zero, and the products $\cos\varphi'\hat{u}_i$ reduce to $\cos^2\varphi'$, $\cos\varphi'\sin\varphi'$ and $\cos\varphi'$, whose even parts are $\cos^2\psi\cos^2\varphi+\sin^2\psi\sin^2\varphi$, $\cos2\psi\sin\varphi\cos\varphi$ and $\cos\psi\cos\varphi$. Collecting the terms, the two columns of Eq.~\eqref{eq:Nazimuth} can be expressed through the four azimuthal integrals
\begin{align}
\cA&=\int_0^{2\pi}\cos\psi\,J_1\,\ud\psi, &
\cB&=\int_0^{2\pi}\cos\psi\,J_2\,\ud\psi,\nonumber\\
\cC&=\int_0^{2\pi}\cos2\psi\,J_2\,\ud\psi, &
\cD+I_0&=\int_0^{2\pi}J_2\,\ud\psi,\nonumber\\
I_0&=\int_0^{2\pi}\cL\,\ud\psi ,
\label{eq:ABCDdef}
\end{align}
where the constant term of the logarithm has been separated out as the function $I_0$. As an example, for $i=x$ the bracket of Eq.~\eqref{eq:Nazimuth} is $\rho\cos\varphi\,J_1-\sin\theta\cos\varphi'\,J_2$, and keeping only the even parts gives
\begin{eqnarray}
\int_0^{2\pi}\cos\varphi'\left[\rho\cos\varphi\,J_1-\sin\theta\cos\varphi'\,J_2\right]\ud\varphi'
=\nonumber\\\rho\cos^2\!\varphi\,\cA-\frac{\sin\theta}{2}\left(\cD+I_0+\cC\cos2\varphi\right) ,
\label{eq:Nxxexample}
\end{eqnarray}
since $\cos^2\!\psi\cos^2\!\varphi+\sin^2\!\psi\sin^2\!\varphi=\frac{1}{2}(1+\cos2\psi\cos2\varphi)$. The other elements follow in the same way. This procedure is equivalent to evaluating the tensor for a field point in the $xz$-plane, $\varphi=0$, and rotating it back to the azimuth of the field point, $\bfN(\rho,\varphi,z)=\mathbb{R}_z(\varphi)\,\bfN(\rho,0,z)\,\mathbb{R}_z(\varphi)^T$, where $\mathbb{R}_z(\varphi)$ is the matrix of the rotation by the angle $\varphi$ about the $z$-axis, which is permitted because the cone is invariant under this rotation. The contribution of the lateral surface to the first two columns thus becomes
\begin{align}
N^\n{Lateral}_{xx}&=-\frac{\sin\theta\cos\theta}{4\pi}\biggl(\rho\cos^2\!\varphi\,\cA\nonumber\\
&\qquad-\frac{\sin\theta}{2}\bigl(\cD+I_0+\cC\cos2\varphi\bigr)\biggr),\nonumber\\
N^\n{Lateral}_{yy}&=-\frac{\sin\theta\cos\theta}{4\pi}\biggl(\rho\sin^2\!\varphi\,\cA\nonumber\\
&\qquad-\frac{\sin\theta}{2}\bigl(\cD+I_0-\cC\cos2\varphi\bigr)\biggr),\nonumber\\
N^\n{Lateral}_{xy}&=N^\n{Lateral}_{yx}\nonumber\\
&=-\frac{\sin\theta\cos\theta}{4\pi}\sin\varphi\cos\varphi\bigl(\rho\cA-\sin\theta\,\cC\bigr),\nonumber\\
N^\n{Lateral}_{zx}&=-\frac{\sin\theta\cos\theta}{4\pi}\cos\varphi\bigl(z\cA-\cos\theta\,\cB\bigr),\nonumber\\
N^\n{Lateral}_{zy}&=-\frac{\sin\theta\cos\theta}{4\pi}\sin\varphi\bigl(z\cA-\cos\theta\,\cB\bigr).
\label{eq:Nlat}
\end{align}

The integrals $\cB$, $\cC$ and $\cD$ contain the logarithm $\cL$ through $J_2$. These logarithms are removed by the identity
\begin{equation}
\frac{\ud\cL}{\ud\psi}=\frac{\partial\cL}{\partial b}\frac{\ud b}{\ud\psi}=-A\sin\psi\,J_1 ,
\label{eq:dL}
\end{equation}
which follows from $\partial\cL/\partial b=\int_0^\ell s\,\ud s/\Delta_s^3$. Integration by parts with the periodic functions $\sin\psi$ and $\sin2\psi$ gives $\int\cos\psi\,\cL\,\ud\psi=A\int\sin^2\!\psi\,J_1\,\ud\psi$ and $\int\cos2\psi\,\cL\,\ud\psi=A\int\sin^2\!\psi\cos\psi\,J_1\,\ud\psi$, so that
\begin{align}
\cB&=\int_0^{2\pi}\bigl[A\sin^2\!\psi\,J_1+2b\cos\psi\,J_1-c^2\cos\psi\,J_0\bigr]\ud\psi,\nonumber\\
\cC&=\int_0^{2\pi}\bigl[A\sin^2\!\psi\cos\psi\,J_1+2b\cos2\psi\,J_1\nonumber\\
&\qquad\quad-c^2\cos2\psi\,J_0\bigr]\ud\psi,\nonumber\\
\cD&=\int_0^{2\pi}\bigl[2bJ_1-c^2J_0\bigr]\ud\psi
\label{eq:BCDreduced}
\end{align}
are integrals of rational functions of $\cos\psi$, multiplied by $1$ or by $1/\Delta_\ell$, over a full period, and hence combinations of complete elliptic integrals and elementary functions. Only the constant term $I_0$ of the logarithm remains, since $\int_0^{2\pi}\cL\,\ud\psi$ has no periodic function to integrate against. 

\subsection{Closed forms}
\label{sec:closed}
The substitution $\psi=\pi-2\eta$, $\cos\psi=2\sin^2\!\eta-1$, brings the three factors that appear in the denominators of Eq.~\eqref{eq:BCDreduced} to the standard form of the elliptic integrals. With the helper quantities
\begin{equation}
\begin{aligned}
\varrho_\pm&=\sqrt{(\rho\pm\Rb)^2+(z-L)^2}, &\qquad m&=\frac{4\rho\Rb}{\varrho_+^2}=1-\frac{\varrho_-^2}{\varrho_+^2},\\
g_\pm&=c\pm z\cos\theta\mp\rho\sin\theta, &\qquad n_\pm&=\mp\frac{2\rho\sin\theta}{g_\pm},\\
e_\pm&=|c\cos\theta\pm z|, &\qquad h_\pm&=A^2-B(B\pm c),\\
k_\pm&=(c\pm B)^2(c\mp B)\pm A^2B ,
\end{aligned}
\label{eq:helpers}
\end{equation}
where $\varrho_+$ and $\varrho_-$ are the distances from the field point to the far and the near point of the rim of the base, we get
\begin{equation}
\begin{aligned}
&\Delta_\ell=\varrho_+\sqrt{1-m\sin^2\!\eta},\quad c\mp b=g_\mp\left(1-n_\mp\sin^2\!\eta\right),\\
&\int_0^{2\pi}\frac{\ud\psi}{c\mp b}=\frac{2\pi}{e_\mp} .
\end{aligned}
\label{eq:subst}
\end{equation}
The parameter $m$ depends only on the position of the field point relative to the rim of the base, since $\Delta_\ell$ is the distance from the field point to the point of the rim at the azimuth $\varphi'$. It is the same parameter as that of the complete elliptic integrals in the potential of a uniformly charged circular ring, and in the field of a circular current loop \citep{Jackson_1999}, of radius $\Rb$ in the plane $z=L$. It vanishes on the axis and is smaller than $1$ everywhere except on the rim itself, where $\varrho_-=0$ and $m=1$. After partial fractions in $\cos\psi$, the integrals of Eq.~\eqref{eq:BCDreduced} are given in terms of the complete elliptic integrals of the first, second and third kind,
\begin{equation}
\begin{aligned}
K(m)&=\int_0^{\pi/2}\frac{\ud\eta}{\sqrt{1-m\sin^2\!\eta}},\\
E(m)&=\int_0^{\pi/2}\sqrt{1-m\sin^2\!\eta}\,\ud\eta,\\
\Pi(n|m)&=\int_0^{\pi/2}\frac{\ud\eta}{(1-n\sin^2\!\eta)\sqrt{1-m\sin^2\!\eta}} ,
\end{aligned}
\label{eq:ellip}
\end{equation}
in the parameter convention of Abramowitz and Stegun \citep{Abramowitz_1964}, which is also that of Matlab and Mathematica, i.e.\ the parameter $m$ is the square of the modulus $k$ of Legendre's form, and the characteristic $n$ enters as $1-n\sin^2\!\eta$. We write $K=K(m)$, $E=E(m)$ and $\Pi_\pm=\Pi(n_\pm|m)$. The characteristic $n_-$ lies in $(0,1]$ and reaches $1$ only on the cone $\vartheta=\theta$ through the lateral surface, with $\vartheta$ being the polar angle of the field point, while $n_+\le0$, so that all three integrals are real and regular away from the lateral surface. The four functions of Eq.~\eqref{eq:ABCDdef} are then
\begin{align}
\cA&=\frac{\pi}{A}\left[\frac{c-B}{e_-}-\frac{c+B}{e_+}\right]\nonumber\\
&\quad-\frac{2}{A\varrho_+}\biggl[\frac{(c-B)(c-\ell)}{g_-}\Pi_-\nonumber\\
&\qquad\quad-\frac{(c+B)(c+\ell)}{g_+}\Pi_++2\ell K\biggr],
\label{eq:calA}\\
\cD&=\pi c\left[\frac{1}{e_-}-\frac{1}{e_+}\right]\nonumber\\
&\quad-\frac{2c}{\varrho_+}\left[\frac{c-\ell}{g_-}\Pi_--\frac{c+\ell}{g_+}\Pi_+\right]-\frac{8\ell}{\varrho_+}K,
\label{eq:calD}\\
\cB&=\frac{\pi}{A}\left[\frac{h_-}{e_-}+\frac{h_+}{e_+}\right]\nonumber\\
&\quad-\frac{2}{A\varrho_+}\left[\frac{h_-(c-\ell)}{g_-}\Pi_-+\frac{h_+(c+\ell)}{g_+}\Pi_+\right]\nonumber\\
&\quad-\frac{4\ell}{Am\varrho_+}\left[(2-m)A+mB\right]K+\frac{8\ell}{m\varrho_+}E,
\label{eq:calB}\\
\cC&=\frac{\pi}{A^2}\left[\frac{k_-}{e_-}-\frac{k_+}{e_+}\right]\nonumber\\
&\quad-\frac{2}{A^2\varrho_+}\left[\frac{k_-(c-\ell)}{g_-}\Pi_--\frac{k_+(c+\ell)}{g_+}\Pi_+\right]\nonumber\\
&\quad-\frac{4}{A^2m^2\varrho_+}\Bigl[8A^2\ell-2m\left(Ac^2+4A^2\ell-AB\ell\right)\nonumber\\
&\qquad\quad+m^2\left(Ac^2+2A^2\ell-AB\ell+(c^2-B^2)\ell\right)\Bigr]K\nonumber\\
&\quad+\frac{8}{Am^2\varrho_+}\left[4A\ell-m\left(c^2+(2A-B)\ell\right)\right]E .
\label{eq:calC}
\end{align}
These expressions contain inverse powers of $A=\rho\sin\theta$ and of $m$, which are both proportional to $\rho$, and they therefore take the form $0/0$ on the axis of the cone. The limit $\rho\to0$ nevertheless exists. On the axis $b=B$ does not depend on $\psi$, so that the integrands of $\cA$, $\cB$ and $\cC$ in Eq.~\eqref{eq:ABCDdef} are a constant times $\cos\psi$ or $\cos2\psi$, and the three functions vanish there, $\cA$ and $\cB$ linearly and $\cC$ quadratically in $\rho$. In the closed forms this appears as a cancellation between the terms. For $\rho\to0$ the parameter $m$ and the characteristics $n_\pm$ vanish, so that $K$, $E$ and $\Pi_\pm$ tend to $\pi/2$, while $e_\pm$ and $g_\pm$ both tend to $c\pm B$, and the terms of each of Eqs.~\eqref{eq:calA}, \eqref{eq:calB} and \eqref{eq:calC} cancel to the order in $\rho$ that compensates the inverse powers of $A$ and $m$. In Eq.~\eqref{eq:calA}, for instance, the first bracket becomes $1-1$ and the second $\frac{\pi}{2}\left[(c-\ell)-(c+\ell)+2\ell\right]$. Only $\cD$ contains no such inverse powers and remains finite,
\begin{equation}
\cD(0,z)=\frac{2\pi}{\sin^2\!\theta}\left[\cos\theta\,\sgn(z)+\frac{\ell\cos2\theta-z\cos\theta}{\rho_1}\right] ,
\label{eq:calDaxis}
\end{equation}
where $\rho_1$ is the distance from the field point on the axis to the rim of the base, given in Eq.~\eqref{eq:I0axis}. In the tensor of Sec.~\ref{sec:tensor} all off-diagonal elements and the dependence on $\varphi$ then disappear, and the on-axis expressions of Sec.~\ref{sec:special} are recovered. 

\subsection{The function $I_0$}
\label{sec:I0}
The remaining function $I_0$ is the integral of $\cL$ over the azimuth. The function $\cL$ of Eq.~\eqref{eq:slant} is, up to the factor $1/4\pi$, the potential at the field point of a generator of the cone, i.e.\ of one of the straight segments from the apex to the rim of the base, when this carries a uniform line charge of unit density. The function $I_0$ is thus the potential of a generator integrated over all azimuths,
\begin{equation}
\begin{aligned}
I_0(\rho,z)&=\int_0^{2\pi}\ln\frac{c+\Delta_\ell+\ell}{c+\Delta_\ell-\ell}\,\ud\psi\\
&=2\int_0^{2\pi}\atanh\frac{\ell}{c+\Delta_\ell}\,\ud\psi ,
\end{aligned}
\label{eq:I0}
\end{equation}
with $\Delta_\ell$ of Eq.~\eqref{eq:Delta}. It depends on the cone only through $\ell$ and $\theta$ and is positive everywhere. Since $\partial\cL/\partial\ell=1/\Delta_\ell$ and $\int_0^{2\pi}\ud\psi/\Delta_\ell=4K(m)/\varrho_+$, it is also the integral of the potential of a ring of unit line charge over the slant length,
\begin{equation}
\begin{aligned}
I_0&=\int_0^{\ell}\frac{4K(m_s)}{\varrho_+(s)}\,\ud s,\\
\varrho_+(s)^2&=(\rho+s\sin\theta)^2+(z-s\cos\theta)^2,\\
m_s&=\frac{4\rho s\sin\theta}{\varrho_+(s)^2} ,
\end{aligned}
\label{eq:I0ring}
\end{equation}
As the area element of the lateral surface is $\ud S'=s\sin\theta\,\ud s\,\ud\varphi'$, this is the integral of $D/(s\sin\theta)$ over the lateral surface. The function $I_0$ is therefore also, up to the factor $1/4\pi$, the potential of the lateral surface when this carries the surface charge density $1/(s\sin\theta)$, which decreases inversely with the distance $s$ from the apex so that every ring of the surface carries the same charge. On the axis both integrands are constant and
\begin{equation}
\begin{aligned}
I_0(0,z)&=2\pi\ln\frac{|z|+\rho_1+\ell}{|z|+\rho_1-\ell}\\
&=2\pi\ln\frac{\ell-z\cos\theta+\rho_1}{|z|-z\cos\theta},\\
\rho_1&=\sqrt{\Rb^2+(L-z)^2} ,
\end{aligned}
\label{eq:I0axis}
\end{equation}
which is the logarithm that appears in the on-axis expressions of Refs.~\citep{Nguyen_2020a,Nguyen_2020b}. Expanding the ring potential in Legendre polynomials, with $\cos\vartheta=z/c$, gives the series
\begin{align}
I_0&=2\pi\ln\frac{4\ell}{(c\pm z)(1\mp\cos\theta)}\nonumber\\
&\quad-2\pi\sum_{l=1}^{\infty}\frac{P_l(\cos\vartheta)P_l(\cos\theta)}{l}\left(\frac{c}{\ell}\right)^l,\quad c<\ell,\nonumber\\
I_0&=2\pi\sum_{l=0}^{\infty}\frac{P_l(\cos\vartheta)P_l(\cos\theta)}{l+1}\left(\frac{\ell}{c}\right)^{l+1},\quad c>\ell ,
\label{eq:I0series}
\end{align}
where the upper signs hold for $\vartheta<\theta$ and the lower for $\vartheta>\theta$. The logarithm in the first series is the limit of $I_0$ for a field point close to the apex, $c\ll\ell$, where the sum vanishes. The series converge geometrically in $c/\ell$ or $\ell/c$ and slowly near the sphere $c=\ell$ through the rim of the base. This means that the numerical evaluation of Eq.~\eqref{eq:I0} is preferable to the above expressions. 


\subsection{The demagnetization tensor}
\label{sec:tensor}
The expressions of Eq.~\eqref{eq:Nlat} simplify when written in terms of the three combinations
\begin{equation}
\begin{aligned}
\cP&=\rho\cA-\sin\theta\,\cC,\qquad
\cQ=z\cA-\cos\theta\,\cB,\\
\cR&=\rho\cA-\sin\theta\left(\cD+I_0\right) ,
\end{aligned}
\label{eq:PQR}
\end{equation}
since $\rho\cos^2\!\varphi\,\cA=\frac{1}{2}\rho\cA(1+\cos2\varphi)$. The point demagnetization tensor is then 
\begin{equation}
\begin{aligned}
\bfN&=\Theta\,\zhat\zhat^T-\frac{\sin\theta\cos\theta}{8\pi}\\
&\quad\times\left(\begin{array}{ccc}
\cR+\cP\cos2\varphi & \cP\sin2\varphi & 2\cQ\cos\varphi\\[1ex]
\cP\sin2\varphi & \cR-\cP\cos2\varphi & 2\cQ\sin\varphi\\[1ex]
2\cQ\cos\varphi & 2\cQ\sin\varphi & -2\cR
\end{array}\right) ,
\end{aligned}
\label{eq:Nmatrix}
\end{equation}
where $\Theta$ is equal to one inside the cone and zero outside. The third column of Eq.~\eqref{eq:Nmatrix}, which is the field of the axially magnetized cone, i.e. $N^\n{Lateral}_{iz}$ in Eqs.~\eqref{eq:latint} and \eqref{eq:baseint}, has been completed from the symmetry $N_{xz}=N_{zx}$, $N_{yz}=N_{zy}$ of the tensor and from the trace rule $N_{zz}=\Theta-N_{xx}-N_{yy}$ \citep{Moskowitz_1966}, and we have verified numerically that it agrees with the direct integration of the charges $M_z$ on the base and $-M_z\sin\theta$ on the lateral surface, see Sec.~\ref{sec:verification}. Equation \eqref{eq:Nmatrix} is thus the complete tensor for an arbitrary direction of the magnetization. The function $I_0$ enters only through $\cR$, i.e.\ through the combination
\begin{equation}
\frac{\sin^2\!\theta\cos\theta}{8\pi}\,I_0\,\operatorname{diag}(1,1,-2) ,
\label{eq:I0part}
\end{equation}
which is traceless and isotropic in the plane perpendicular to the axis. All other parts of the tensor are given in closed form by Eqs.~\eqref{eq:calA} to \eqref{eq:calC}.

\subsection{The demagnetization vector}
\label{sec:vector}
The magnetic scalar potential follows from the same integration along the generators, now of the kernel itself,
\begin{equation}
\int_0^\ell\frac{s\,\ud s}{\Delta_s}=\Delta_\ell-c+b\cL ,
\label{eq:slantphi}
\end{equation}
For the lateral surface the two transverse components of Eq.~\eqref{eq:Nphi} are
\begin{equation}
\begin{aligned}
&\left(\begin{array}{c}\Psi^\n{Lateral}_{x}\\[0.5ex] \Psi^\n{Lateral}_{y}\end{array}\right)=\frac{\sin\theta\cos\theta}{4\pi}\\
&\quad\times\left[\cU+AB\,\cV+\frac{A^2}{2}\,\cW+\frac{A}{2}\,I_0\right]\left(\begin{array}{c}\cos\varphi\\ \sin\varphi\end{array}\right) ,
\end{aligned}
\label{eq:Nphixy}
\end{equation}
with three azimuthal integrals
\begin{align}
\cU&=\int_0^{2\pi}\cos\psi\,\Delta_\ell\,\ud\psi\nonumber\\
&=\frac{4\varrho_+}{3m}\left[2(1-m)K+(m-2)E\right],\nonumber\\
\cV&=\int_0^{2\pi}\sin^2\!\psi\,J_1\,\ud\psi\nonumber\\
&=\frac{\pi}{A^2}\left[2c-e_--e_+\right]\nonumber\\
&\quad+\frac{2}{A^2\varrho_+}\bigl[(c-B-A)(c-\ell)\Pi_-\nonumber\\
&\qquad\quad+(c+B+A)(c+\ell)\Pi_+\bigr]\nonumber\\
&\quad-\frac{4}{A^2m\varrho_+}\left[m\left(c^2+(A+B)\ell\right)-2A\ell\right]K\nonumber\\
&\quad-\frac{8\ell}{Am\varrho_+}E,\nonumber\\
\cW&=\int_0^{2\pi}\sin^2\!\psi\cos\psi\,J_1\,\ud\psi\nonumber\\
&=\frac{\pi}{A^3}\left[(c+B)e_+-(c-B)e_--4Bc\right]\nonumber\\
&\quad+\frac{2}{A^3\varrho_+}\bigl[(c-B)(c-B-A)(c-\ell)\Pi_-\nonumber\\
&\qquad\quad-(c+B)(c+B+A)(c+\ell)\Pi_+\bigr]\nonumber\\
&\quad-\frac{4}{3A^3m^2\varrho_+}\bigl[-8A^2\ell\nonumber\\
&\qquad\quad+2m\left(3Ac^2+4A^2\ell+3AB\ell\right)\nonumber\\
&\qquad\quad-3m^2\left(Ac^2+2Bc^2+AB\ell+B^2\ell+c^2\ell\right)\bigr]K\nonumber\\
&\quad+\frac{8}{3A^2m^2\varrho_+}\left[m\left(3c^2+(2A+3B)\ell\right)-4A\ell\right]E .
\label{eq:UVW}
\end{align}
For the axial component the lateral surface carries the charge $-M_z\sin\theta$ on the area element $s\sin\theta\,\ud s\,\ud\varphi'$ and thus gives
\begin{equation}
\begin{aligned}
&\Psi^\n{Lateral}_{z}=-\frac{\sin^2\!\theta}{4\pi}\int_0^{2\pi}\left(\Delta_\ell-c+b\cL\right)\ud\psi\\
&\quad=-\frac{\sin^2\!\theta}{4\pi}\left(4\varrho_+E-2\pi c+BI_0+A^2\cV\right) ,
\end{aligned}
\label{eq:Nphiz1}
\end{equation}
where $\int_0^{2\pi}\Delta_\ell\,\ud\psi=4\varrho_+E$ and the identity of Eq. \eqref{eq:dL} have been used. The base contributes the potential of a uniformly charged disk of radius $\Rb$ in the plane $z'=L$, whose rim is that of the cone, so that $\varrho_\pm$ and $m$ are those of Eq.~\eqref{eq:helpers}. Integrating first over the radius of the disk and then over the azimuth, in the same way as in Sec.~\ref{sec:closed}, gives
\begin{align}
&\Psi^\n{Base}_{z}=\frac{1}{4\pi}\int_0^{2\pi}\!\!\int_0^{\Rb}D\,\rho'\,\ud\rho'\,\ud\varphi'\Big|_{z'=L}\nonumber\\
&\quad=\frac{1}{4\pi}\biggl\{2\varrho_+E+\frac{2(\Rb^2-\varrho_0^2)}{\varrho_+}K-2\pi|z-L|\nonumber\\
&\qquad+\frac{2}{\varrho_+}\Bigl[(\varrho_0-\Rb)(\varrho_0-\rho)\Pi_\n{b}^+\nonumber\\
&\qquad\qquad+(\varrho_0+\Rb)(\varrho_0+\rho)\Pi_\n{b}^-\Bigr]\biggr\},
\label{eq:disk}
\end{align}
with
\begin{equation}
\varrho_0=\sqrt{\rho^2+(z-L)^2},\qquad\qquad \Pi_\n{b}^\pm=\Pi\!\left(\frac{2\rho}{\rho\pm\varrho_0}\,\Big|\,m\right) ,
\label{eq:diskhelpers}
\end{equation}
where $\varrho_0$ is the distance from the field point to the centre of the base. The characteristics of $\Pi_\n{b}^\pm$ are those of the end surfaces of the cylindrical tile \citep{Slanovc_2022,Nielsen_2026}. On the axis Eq.~\eqref{eq:disk} reduces to $2\pi[\sqrt{\Rb^2+(z-L)^2}-|z-L|]/4\pi$. The demagnetization vector is then
\begin{equation}
\pmb{\Psi}=\left(\begin{array}{c}\Psi^\n{Lateral}_{x}\\[0.5ex] \Psi^\n{Lateral}_{y}\\[0.5ex] \Psi^\n{Lateral}_{z}+\Psi^\n{Base}_{z}\end{array}\right) ,
\label{eq:Nphi_total}
\end{equation}
and it thus has the closed form with the same elliptic integrals as the point demagnetization tensor up to the one function $I_0$, which enters the transverse components through $AI_0/2$ and the axial component through $-B\sin^2\!\theta\,I_0$. We note that the tensor of Eq.~\eqref{eq:Nmatrix} is the gradient of Eq.~\eqref{eq:Nphi_total} according to Eq.~\eqref{eq:Ndef}, which provides an independent check of both, see Sec.~\ref{sec:verification}.

\subsection{Special cases and singular points}
\label{sec:special}
On the axis, $\rho=0$, the elliptic integrals reduce to $\pi/2$ and the rotation of Sec.~\ref{sec:generators} is undefined, and the tensor is diagonal with $N_{yy}=N_{xx}$ and
\begin{align}
&N_{xx}(0,0,z)=\frac{\cos\theta}{4}\biggl[\sin^2\!\theta\ln\frac{\ell-z\cos\theta+\rho_1}{|z|-z\cos\theta}\nonumber\\
&\qquad+\frac{\ell(2\cos^2\!\theta-1)-z\cos\theta}{\rho_1}+\cos\theta\,\sgn(z)\biggr],\nonumber\\
&N_{zz}(0,0,z)=\frac{1}{2}\left[\sgn(L-z)-\frac{L-z}{\rho_1}\right]\nonumber\\
&\qquad+\frac{\sin^2\!\theta}{2}\biggl[\frac{2L-z}{\rho_1}+\sgn(z)\nonumber\\
&\qquad\qquad-\cos\theta\ln\frac{\ell-z\cos\theta+\rho_1}{|z|-z\cos\theta}\biggr],
\label{eq:axis}
\end{align}
with $\rho_1$ of Eq.~\eqref{eq:I0axis}, which are obtained by direct integration of the surface charges along the axis and agree with Refs.~\citep{Nguyen_2020a,Nguyen_2020b}. The logarithm in Eq.~\eqref{eq:axis} is $I_0(0,z)/2\pi$ with the coefficient of Eq.~\eqref{eq:I0part}. 

Numerically we find that the cancellation of the inverse powers of $A$ and $m$ in Eqs.~\eqref{eq:calA} to \eqref{eq:calC} and \eqref{eq:UVW} is harmless in double precision down to $\rho/\ell\approx10^{-3}$, where the closed form still agrees with Eq.~\eqref{eq:axis} to $3\times10^{-8}$, which is the $\rho^2$ correction itself, and Eq.~\eqref{eq:axis} should be used below that.

The individual terms of Eqs.~\eqref{eq:calA} to \eqref{eq:calC} and \eqref{eq:UVW} are singular on the cone $\vartheta=\theta$ beyond the base, where $e_-=0$ and $n_-=1$, and on the mirror cone $\vartheta=\pi-\theta$ below the apex, where $e_+=0$ and $g_+=0$. On these two surfaces the field point lies on the line that contains a generator, $d^2=0$ for one value of $\psi$, and the singularities cancel between the terms, as the field itself is regular there. For field points in the plane of the base, $z=L$, the characteristic of $\Pi_\n{b}^-$ in Eq.~\eqref{eq:disk} diverges, as for the end surfaces of the cylindrical tile, and $\Pi_\n{b}^-$ must be replaced by its limit. The solution as given therefore cannot be evaluated directly on the axis, on the two cones through the apex, on the lateral surface, on the rim of the base, where $m=1$, and, for the axial component of the demagnetization vector, in the plane of the base, and requires the limits stated above there. The field on the lateral surface is discontinuous, as for every charged surface.

The expressions above are for a cone with its apex at the origin. For a cone that stands on its base, with the centre of the base at the origin and the apex at $(0,0,L)$, the field point $(x,y,z)$ is the mirror image in the plane $z=L/2$ of the point $(x,y,L-z)$ of the cone considered here. The demagnetization tensor and the demagnetization vector of the cone standing on its base are therefore
\begin{equation}
\begin{aligned}
N^\n{up}_{ij}(x,y,z)&=\epsilon_i\epsilon_j\,N_{ij}(x,y,L-z),\\
\Psi^\n{up}_{j}(x,y,z)&=\epsilon_j\,\Psi_{j}(x,y,L-z),\\
\epsilon_x&=\epsilon_y=1,\qquad \epsilon_z=-1 ,
\end{aligned}
\label{eq:upright}
\end{equation}
with no summation over repeated indices, i.e. only the elements $N_{xz}=N_{zx}$ and $N_{yz}=N_{zy}$ of the tensor and the component $\Psi_z$ of the vector change sign.

\subsection{The truncated cone and the annular cone}
\label{sec:truncated}
A truncated cone with the slant lengths $\ell_1<\ell_2$ is the difference of two cones with the same apex and the same half-opening angle, and as both are homogeneously magnetized with the same magnetization, its tensor and its demagnetization vector are the differences of Eqs.~\eqref{eq:Nmatrix} and \eqref{eq:Nphi_total} evaluated for $\ell_2$ and for $\ell_1$. In particular the function $I_0$ of the truncated cone is $\int_{\ell_1}^{\ell_2}4K(m_s)\,\ud s/\varrho_+(s)$ by Eq.~\eqref{eq:I0ring}. On the axis the field of the truncated cone follows from the difference of Eq.~\eqref{eq:axis} for the two cones. At the virtual apex, $z=0$, which lies outside the truncated cone, the base terms and the terms $2L/\rho_1$ of the two cones cancel, as do the terms with $\sgn(z)$, and the logarithms leave
\begin{equation}
\begin{aligned}
H_z(0,0,0)&=\frac{M_z}{2}\sin^2\!\theta\cos\theta\,\ln\frac{\ell_2}{\ell_1},\\
\frac{\partial H_z}{\partial z}\Big|_{z=0}&=\frac{3M_z}{2}\sin^2\!\theta\cos^2\!\theta\left(\frac{1}{\ell_1}-\frac{1}{\ell_2}\right),
\end{aligned}
\label{eq:apex}
\end{equation}
where $\ell_2/\ell_1$ is also the ratio of the radii of the two bases. The apex field is the result of Batishchev and Hyde \citep{Batishchev_2023}, and it is largest for $\tan^2\!\theta=2$. For bases at fixed heights $z_1=\ell_1\cos\theta$ and $z_2=\ell_2\cos\theta$ the gradient becomes $\frac{3M_z}{2}\sin^2\!\theta\cos^3\!\theta\,(1/z_1-1/z_2)$ and is largest for $\tan^2\!\theta=2/3$, also in agreement with Ref.~\citep{Batishchev_2023}. In the same way an annular cone, i.e.\ a cone with a coaxial conical bore, is the difference of two cones with the same apex and different half-opening angles. A conical sector, i.e.\ a cone cut by two planes through the axis, would require the incomplete elliptic integrals and is not treated here.

\section{The magnetometric tensor}
\label{sec:magnetometric}
The magnetometric demagnetization tensor is the average of the point tensor over the volume $V$ of the body,
\begin{equation}
\bfD=\frac{1}{V}\int_V\bfN\,\ud V ,
\label{eq:Ddef_m}
\end{equation}
which gives the average field in the cone, $\langle\bfH\rangle=-\bfD\cdot\bfM$, and the magnetostatic self-energy $E=\frac{1}{2}\mu_0V\,\bfM\cdot\bfD\cdot\bfM$ \citep{Moskowitz_1966}. As $N_{ij}=\partial_i\Psi_j$ by Eq.~\eqref{eq:Ndef}, the divergence theorem turns the volume integral into a surface integral of the demagnetization vector, and inserting Eq.~\eqref{eq:Nphi} gives a double integral over all pairs of points on the surface of the body,
\begin{equation}
D_{ij}=\frac{1}{V}\oint_{S}n_i\,\Psi_j\,\ud S=\frac{1}{4\pi V}\oint_{S}\oint_{S'}n_i\,n'_j\,D\,\ud S'\,\ud S .
\label{eq:Dface}
\end{equation}
For the cone the tensor is diagonal with $D_{xx}=D_{yy}$ by symmetry, and since the trace of $\bfN$ is $\Theta$, the trace of $\bfD$ is one, so that $D_{zz}=1-2D_{xx}$. It is therefore sufficient to compute the transverse factor, and to this only the lateral surface contributes, because the normal of the base has no transverse component. With the normal $\nhat'=(\cos\theta\cos\varphi',\cos\theta\sin\varphi',-\sin\theta)$ of the lateral surface,
\begin{equation}
D_{xx}=\frac{\cos^2\!\theta}{4\pi V}\int_{S_1}\int_{S_1'}\cos\varphi\cos\varphi'\,D\,\ud S'\,\ud S ,
\label{eq:Dxx_def}
\end{equation}
where both points are on the lateral surface. This integral is also the field energy of the azimuthal current sheet $M\cos\theta$ on the lateral surface, which is the equivalent current of the axially magnetized cone considered by Smith et al.~\citep{Smith_2021}, in agreement with the relation $E=\frac{1}{2}\mu_0VM^2(1-D_{zz})$ between the energy of the field $\bfB$ of the equivalent currents and the demagnetizing energy.

For two points $(s,\varphi)$ and $(s',\varphi')$ on the lateral surface, with $\ud S=s\sin\theta\,\ud s\,\ud\varphi$, the distance is
\begin{equation}
\begin{aligned}
\|\bfr-\bfr'\|^2&=s^2+s'^2-2ss'w,\\
w&=\cos^2\!\theta+\sin^2\!\theta\cos\psi ,
\end{aligned}
\label{eq:wdef}
\end{equation}
where $w$ is the cosine of the angle between the two generators and $\psi=\varphi'-\varphi$ as before. The integration over the two slant coordinates is elementary. By the symmetry in $s$ and $s'$ and the substitution $s'=st$,
\begin{equation}
\begin{aligned}
&\int_0^\ell\!\!\int_0^\ell\frac{ss'\,\ud s\,\ud s'}{\sqrt{s^2+s'^2-2ss'w}}\\
&\quad=2\int_0^\ell s^2\,\ud s\int_0^1\frac{t\,\ud t}{\sqrt{1-2wt+t^2}}=\frac{2\ell^3}{3}\,h(u) ,
\end{aligned}
\label{eq:hdef}
\end{equation}
with
\begin{equation}
\begin{aligned}
h(u)&=2u-1+(1-2u^2)\ln\frac{1+u}{u},\\
u&=\sqrt{\frac{1-w}{2}}=\sin\theta\,\Big|\sin\frac{\psi}{2}\Big| ,
\end{aligned}
\label{eq:h}
\end{equation}
where $2\ell u$ is the distance between the two rim points. As the integrand depends on the azimuths only through their difference, the two azimuthal integrations reduce to one,
\begin{equation}
\begin{aligned}
&\int_0^{2\pi}\!\!\int_0^{2\pi}\cos\varphi\cos\varphi'\,F(\varphi'-\varphi)\,\ud\varphi\,\ud\varphi'\\
&\qquad=\pi\int_0^{2\pi}\cos\psi\,F(\psi)\,\ud\psi ,
\end{aligned}
\label{eq:azred}
\end{equation}
and with the volume of the cone, $V=\frac{1}{3}\pi\Rb^2L=\frac{1}{3}\pi\ell^3\sin^2\!\theta\cos\theta$, and the half azimuth $\chi=\psi/2$ as the integration variable, we obtain
\begin{equation}
\begin{aligned}
D_{xx}&=\frac{\cos\theta}{2\pi}\int_0^{2\pi}\cos\psi\;h\Big(\sin\theta\,\Big|\sin\frac{\psi}{2}\Big|\Big)\,\ud\psi\\
&=\frac{2\cos\theta}{\pi}\int_0^{\pi/2}\cos2\chi\;h(\sin\theta\sin\chi)\,\ud\chi .
\end{aligned}
\label{eq:Dxx_int}
\end{equation}
The slant length has dropped out, as it must, since the magnetometric tensor of a body that is determined up to its size by a single parameter can only depend on that parameter.

\begin{figure}[!ht]
\centering
\includegraphics[width=\columnwidth]{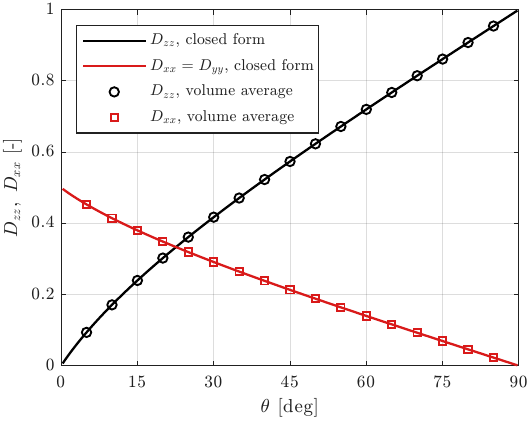}
\caption{The magnetometric demagnetization factors of the cone, Eq.~\eqref{eq:Dcone}, as functions of the half-opening angle $\theta$, together with the volume average of the point tensor of Eq.~\eqref{eq:Nmatrix} computed numerically.}
\label{fig:factors}
\end{figure}

Inserting Eq.~\eqref{eq:h}, the integrand of Eq.~\eqref{eq:Dxx_int} consists of even powers of $\sin\chi$ multiplied by $\ln(1+\kappa\sin\chi)$ and by $\ln(\kappa\sin\chi)$, with $\kappa=\sin\theta$. The integrals of the second kind are elementary, $\int_0^{\pi/2}\ln\sin\chi\,\ud\chi=-\frac{\pi}{2}\ln2$, $\int_0^{\pi/2}\sin^2\!\chi\ln\sin\chi\,\ud\chi=\frac{\pi}{8}(1-2\ln2)$ and $\int_0^{\pi/2}\sin^4\!\chi\ln\sin\chi\,\ud\chi=\frac{\pi}{64}(7-12\ln2)$. For the integrals of the first kind,
\begin{equation}
L_n(\kappa)=\int_0^{\pi/2}\sin^{2n}\!\chi\,\ln(1+\kappa\sin\chi)\,\ud\chi,\qquad n=0,1,2 ,
\label{eq:Lndef}
\end{equation}
the derivative with respect to $\kappa$ is the integral of a rational function of $\sin\chi$, which is elementary, as $\int_0^{\pi/2}\ud\chi/(1+\kappa\sin\chi)=(\pi/2-\beta)/\cos\beta$ with $\kappa=\sin\beta$. Integrating back from $\kappa=0$ with the substitution $\kappa=\sin\beta'$ gives elementary functions of $\beta$ and the single non-elementary integral
\begin{equation}
\begin{aligned}
\int_0^\beta\frac{t\,\ud t}{\sin t}&=2\,\mathrm{Ti}_2\!\Big(\tan\frac{\beta}{2}\Big),\\
\mathrm{Ti}_2(x)&=\int_0^x\frac{\atan t}{t}\,\ud t=\sum_{k=0}^\infty\frac{(-1)^k x^{2k+1}}{(2k+1)^2} ,
\end{aligned}
\label{eq:Ti2}
\end{equation}
where $\mathrm{Ti}_2$ is the inverse tangent integral \citep{Lewin_1981}, which is related to the Clausen function $\mathrm{Cl}_2(\omega)=-\int_0^\omega\ln|2\sin(t/2)|\,\ud t$ by $\mathrm{Ti}_2(\tan\omega)=\omega\ln\tan\omega+\frac{1}{2}\left[\mathrm{Cl}_2(2\omega)+\mathrm{Cl}_2(\pi-2\omega)\right]$ and takes the value $\mathrm{Ti}_2(1)=G=0.915966$, Catalan's constant. Here $\kappa=\sin\theta$, so $\beta=\theta$, and writing $T=\mathrm{Ti}_2(\tan\frac{\theta}{2})$ we find
\begin{align}
L_0&=\frac{\pi}{2}\ln\frac{1+\cos\theta}{2}+2T ,\nonumber\\
L_1&=\frac{\pi}{4}\ln\frac{1+\cos\theta}{2}+\frac{\sin\theta-\theta\cos\theta}{2\sin^2\!\theta}\nonumber\\
&\quad-\frac{\pi}{4(1+\cos\theta)}+\frac{\pi}{8}+T ,\nonumber\\
L_2&=\frac{3\pi}{16}\ln\frac{1+\cos\theta}{2}+\frac{7\pi}{64}+\frac{7}{24\sin\theta}-\frac{\pi}{8\sin^2\!\theta}\nonumber\\
&\quad+\frac{1}{4\sin^3\!\theta}-\frac{\pi}{8\sin^4\!\theta}+\frac{3}{4}T\nonumber\\
&\quad+\Big(\frac{\pi}{2}-\theta\Big)\cos\theta\left(\frac{1}{4\sin^4\!\theta}+\frac{3}{8\sin^2\!\theta}\right) .
\label{eq:Ln}
\end{align}
When the terms are collected in Eq.~\eqref{eq:Dxx_int} almost everything cancels, and the magnetometric tensor of the cone becomes
\begin{equation}
\begin{aligned}
D_{zz}&=\sin^2\!\theta+\frac{2\cos\theta}{\pi}\left(\sin\theta+\theta\cos\theta\right)\\
&\quad-\cos\theta\sin^2\!\theta\left[\ln\cot\frac{\theta}{2}+\frac{4}{\pi}\,\mathrm{Ti}_2\!\Big(\tan\frac{\theta}{2}\Big)\right],
\end{aligned}
\label{eq:Dcone}
\end{equation}
with the transverse factors $D_{xx}=D_{yy}=\frac{1}{2}(1-D_{zz})$,
or, in terms of the Clausen function,
\begin{equation}
\begin{aligned}
D_{zz}&=\sin^2\!\theta+\frac{2\cos\theta}{\pi}\biggl[\sin\theta+\theta\cos\theta\\
&\quad-\sin^2\!\theta\biggl(\Big(\frac{\pi}{2}-\theta\Big)\ln\cot\frac{\theta}{2}\\
&\qquad\quad+\mathrm{Cl}_2(\theta)+\mathrm{Cl}_2(\pi-\theta)\biggr)\biggr] .
\end{aligned}
\label{eq:DconeCl}
\end{equation}
The two factors are shown in Fig.~\ref{fig:factors} as functions of the half-opening angle. In the limit of a flat cone, $\theta\to\pi/2$, the axial factor tends to one as $1-D_{zz}\simeq\frac{2}{\pi}(2G-1)(\frac{\pi}{2}-\theta)$, with the slope set by Catalan's constant, and in the limit of a needle, $\theta\to0$, it vanishes as $D_{zz}\simeq\frac{4}{\pi}\theta+\theta^2\big(1-\ln\frac{2}{\theta}\big)$, i.e.\ linearly in the ratio $\Rb/L$ of the base radius to the height, since the volume average is dominated by the region near the base. The cone is magnetostatically isotropic, $\bfD=\frac{1}{3}\mathbb{I}$, for $\theta=0.39423$~rad, i.e.\ $\theta=22.59^\circ$ or $L/2\Rb=1.2019$, which may be compared with the isotropic aspect ratio $L/2R=0.9065$ of the cylinder \citep{Sakurai_2007}. For the cone with $\theta=\pi/4$, where the height equals the base radius, $D_{zz}=0.57359$.

As the trace of $\bfD$ also determines the sum of the remaining face pairs of Eq.~\eqref{eq:Dface}, the mutual Newtonian energy of the base and the lateral surface follows as well. The self-energy of the base disk is $\int_{S_2}\int_{S_2'}D\,\ud S'\ud S=\frac{16\pi}{3}\Rb^3$, the self-energy of the lateral surface is, by Eq.~\eqref{eq:hdef} with $\int_0^{2\pi}\!\int_0^{2\pi}F(\varphi'-\varphi)\,\ud\varphi\,\ud\varphi'=2\pi\int_0^{2\pi}F(\psi)\,\ud\psi$,
\begin{align}
&\int_{S_1}\int_{S_1'}D\,\ud S'\ud S=\frac{16\pi}{3}\ell^3\sin^2\!\theta\,\mathcal{H}(\theta),\nonumber\\
&\mathcal{H}(\theta)=\int_0^{\pi/2}h(\sin\theta\sin\chi)\,\ud\chi\nonumber\\
&\quad=\sin\theta+\theta\cos\theta-\frac{\pi}{2}\cos\theta\nonumber\\
&\qquad+\cos^2\!\theta\left[\frac{\pi}{2}\ln\cot\frac{\theta}{2}+2\,\mathrm{Ti}_2\!\Big(\tan\frac{\theta}{2}\Big)\right] ,
\label{eq:Hlat}
\end{align}
and since, by Eq.~\eqref{eq:Dface},
\begin{equation*}
\begin{aligned}
D_{zz}=\frac{1}{4\pi V}\Bigl[&\int_{S_2}\!\int_{S_2'}D-2\sin\theta\int_{S_2}\!\int_{S_1'}D\\
&+\sin^2\!\theta\int_{S_1}\!\int_{S_1'}D\Bigr] ,
\end{aligned}
\end{equation*}
the mutual energy of the disk and the conical surface on its rim is
\begin{equation}
\begin{aligned}
&\int_{S_2}\int_{S_1'}D\,\ud S'\ud S=\frac{2\pi}{3}\ell^3\bigl[4\sin^2\!\theta\\
&\qquad+4\sin^3\!\theta\,\mathcal{H}(\theta)-\pi\sin\theta\cos\theta\,D_{zz}\bigr] .
\end{aligned}
\label{eq:mutual}
\end{equation}
All three face-pair integrals of the cone are thus known in closed form, although the potential of the base at a point of the lateral surface contains the function $I_0$.

\section{Verification}
\label{sec:verification}
\begin{figure}[!t]
\centering
\includegraphics[width=\columnwidth]{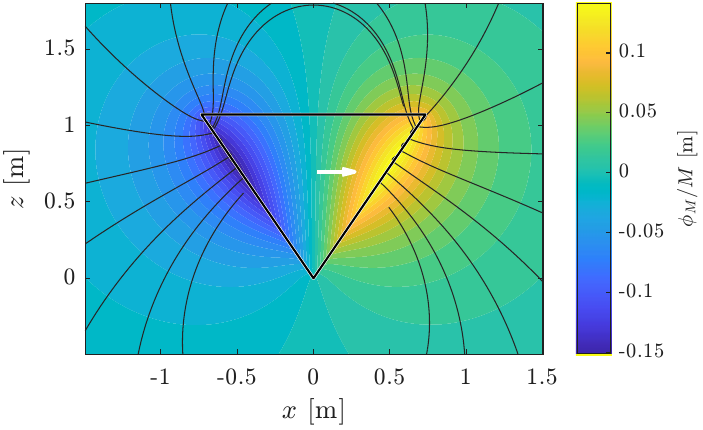}
\caption{The magnetic scalar potential, in units of the magnetization, and the field lines of $\bfH$ in the $xz$-plane, $y=0$, which contains the axis of the cone, for the cone with $\theta=0.6$ and $\ell=1.3$~m magnetized along $x$, as indicated by the arrow. The black line is the cross-section of the cone with this plane, which is the triangle formed by two opposite generators and a diameter of the base, with the apex at the origin and the base at $z=L$.}
\label{fig:fieldmap}
\end{figure}

The expressions derived above are verified against direct numerical integration of Eq.~\eqref{eq:Ndef} and Eq.~\eqref{eq:Nphi} over the two surfaces of the cone, using adaptive two-dimensional quadrature with a tolerance of $10^{-12}$. We consider a cone with $\theta=0.6$ rad and $\ell=1.3$~m, i.e.\ $L=1.073$~m and $\Rb=0.734$~m. Figure~\ref{fig:fieldmap} shows the magnetic scalar potential and the field lines in the $xz$-plane for a magnetization along $x$, computed from Eqs.~\eqref{eq:Nmatrix} and \eqref{eq:Nphi_total}.

Table~\ref{tab:check} compares the tensor at eight field points inside the cone, beside it, above the base and below the apex. All nine components agree with the direct integration to $5\times10^{-13}$, which is the precision of the latter, the tensor is symmetric and its trace is $1$ inside and $0$ outside the cone to $10^{-12}$, and the two representations of $I_0$, Eqs.~\eqref{eq:I0} and \eqref{eq:I0ring}, agree to all printed digits. The third column, obtained in Sec.~\ref{sec:tensor} from the symmetry and the trace of the tensor, was compared with the direct integration of the charges of the axially magnetized cone over the base and the lateral surface at the same points, with the same agreement. The three components of the demagnetization vector agree with the direct potential integrals to $5\times10^{-14}$, and all nine components of the tensor agree with central differences of the demagnetization vector with the step $10^{-4}$~m to $2\times10^{-8}$, which is the truncation error of the differences.

\begin{table*}[!ht]
\caption{The tensor at eight field points for $\theta=0.6$ and $\ell=1.3$~m. The column $\Theta$ marks points inside the cone, $\max|\delta\bfN|$ is the largest deviation of any of the nine components from the direct surface integration and $\max|\delta\pmb{\Psi}|$ that of the three components of the demagnetization vector.}
\label{tab:check}
\centering
\begin{tabular}{rrrcrrrrr}
\toprule
$x$ (m) & $y$ (m) & $z$ (m) & $\Theta$ & $N_{xx}$ & $N_{yy}$ & $N_{zz}$ & $\max|\delta\bfN|$ & $\max|\delta\pmb{\Psi}|$\\
\midrule
 0.30 & 0.20 & 0.50 & 0 & $-0.081312$ & $0.157835$ & $-0.076523$ & $4\times10^{-14}$ & $9\times10^{-15}$\\
 0.10 & $-0.40$ & 1.00 & 1 & $0.193479$ & $0.185296$ & $0.621225$ & $2\times10^{-13}$ & $6\times10^{-15}$\\
 0.90 & 0.50 & 0.40 & 0 & $-0.030031$ & $0.014578$ & $0.015453$ & $6\times10^{-15}$ & $9\times10^{-15}$\\
 0.20 & 0.10 & $-0.60$ & 0 & $0.015634$ & $0.016490$ & $-0.032125$ & $2\times10^{-14}$ & $3\times10^{-15}$\\
 1.50 & 0.30 & 0.90 & 0 & $-0.026713$ & $0.012157$ & $0.014556$ & $4\times10^{-15}$ & $3\times10^{-14}$\\
 0.40 & 0.30 & 1.30 & 0 & $0.064715$ & $0.075378$ & $-0.140094$ & $1\times10^{-14}$ & $2\times10^{-14}$\\
$-0.50$ & 0.10 & 0.20 & 0 & $-0.033026$ & $0.086494$ & $-0.053468$ & $1\times10^{-14}$ & $2\times10^{-15}$\\
 0.05 & 0.02 & 0.90 & 1 & $0.235420$ & $0.235458$ & $0.529122$ & $5\times10^{-13}$ & $5\times10^{-14}$\\
\bottomrule
\end{tabular}
\end{table*}

\begin{figure*}[!t]
\centering
\includegraphics[width=\textwidth]{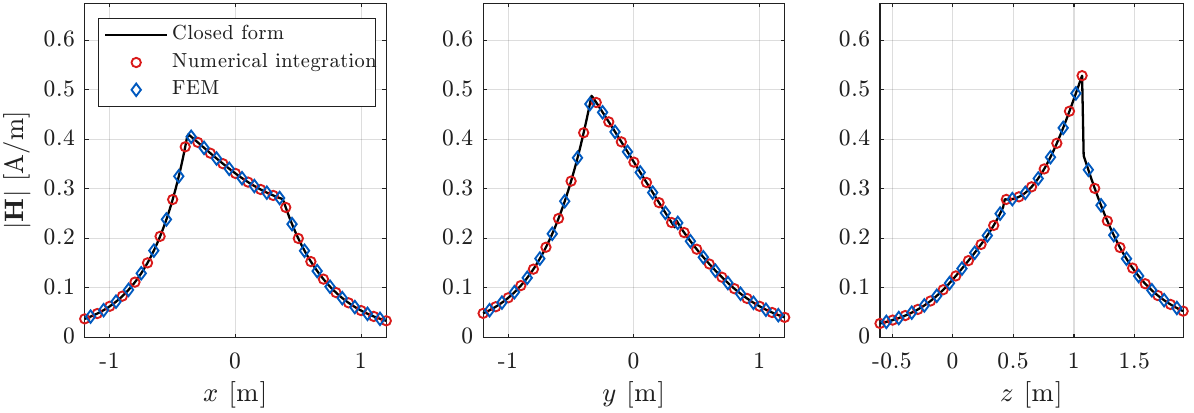}
\caption{The norm of the magnetic field as calculated using the components of the tensor given above, by direct numerical integration of the surface charges and using the FEM framework Comsol, for the cone and the magnetization specified in the text, along each of the Cartesian axes through the point $(x,y,z)=(0.25,0.15,0.6)$~m.}
\label{fig:validation}
\end{figure*}

For the same cone and the magnetization $\bfM=[0.32,0.74,0.89]$~A/m, the norm of the magnetic field is calculated along each of the Cartesian axes through the point $(x,y,z)=(0.25,0.15,0.6)$~m inside the cone and is shown in Fig.~\ref{fig:validation}, together with the direct numerical integration. Additionally, the magnetic field has been computed using the finite element framework Comsol \citep{Comsol}, which solves the magnetic scalar potential equation
\begin{equation}
-\nabla\cdot(\mu_0\nabla\phi_M)=-\nabla\cdot(\mu_0\bfM)
\end{equation}
for the given magnetization, after which the magnetic field is calculated as $\bfH=-\nabla\phi_M$. For the FEM model a highly refined finite element mesh is used to ensure a high precision in the computed field, and a sufficiently large volume is modelled to make the boundary conditions not influence the generated magnetic field. This is also shown on the figure. As can be seen from the figure, there is an excellent agreement between the closed form, the direct integration and the finite element model, also across the lateral surface of the cone, where the norm of the field is discontinuous. The finite element result agrees with the closed form to within $0.5\%$ at all points shown, with a median deviation below $0.1\%$. 

The magnetometric tensor of Eq.~\eqref{eq:Dcone} has been verified in three ways at nine half-opening angles between $0.2$ and $1.4$~rad. The azimuthal integral of Eq.~\eqref{eq:Dxx_int} evaluated numerically reproduces the closed form to $2\times10^{-15}$. The direct evaluation of the three face-pair integrals of Eq.~\eqref{eq:Dface}, with the integral over the base and the lateral surface computed by two-dimensional quadrature, agrees with the closed form to $6\times10^{-12}$ and with Eq.~\eqref{eq:mutual} to the same precision. Finally, the volume average of $N_{zz}$ of Eq.~\eqref{eq:Nmatrix} over the meridian section of the cone, computed with a $64\times64$ Gauss--Legendre rule, agrees with the closed form to $7\times10^{-8}$ or better, the difference being the discretization error of the rule, which decreases when the number of nodes is increased. 

\section{Conclusion}
We have calculated the magnetic field, formulated as the point demagnetization tensor, and the magnetic scalar potential, formulated as the demagnetization vector, for a homogeneously magnetized right circular cone at any point inside or outside the cone, and for the truncated and the annular cone by superposition. All nine components of the tensor and the three components of the vector are given in terms of elementary functions, complete elliptic integrals of the first, second and third kind, and a single scalar function of the field point, the azimuthal average of the potential of a generator of the cone, which cannot be reduced to elliptic integrals and is given as a one-dimensional integral and as a series. The magnetometric demagnetization tensor of the cone has been obtained in closed form as a function of the half-opening angle alone, in terms of elementary functions and the inverse tangent integral. There is an excellent agreement between the field calculated using the tensor and the field computed using direct numerical integration, and between the closed-form magnetometric factors and the volume average of the point tensor.

\section*{Data statement}
The derived analytical expressions for the demagnetization tensor, the demagnetization vector and the magnetometric tensor of the cone are implemented in Matlab together with the verification examples of Sec.~\ref{sec:verification}. These can be found as part of the MagTense open source simulation framework \cite{MagTense,Bjork_2021}.

\section*{Declaration of generative AI and AI-assisted technologies in the manuscript preparation process}
During the preparation of this work the authors used Anthropic Claude, model Fable 5.1 to suggest analytical solutions to the surface integrals. The model was also used to more quickly reformat equations, substitute expressions and simplify the mathematical expressions. Finally, the model was used to provide text on the analytical equations and the validation, to ensure these were mathematically correctly written. After using this model, the authors reviewed and edited the content as needed and takes full responsibility for the content of the published article.

\section*{Acknowledgements}
This work was supported by the Carlsberg Foundation Semper Ardens Advance project CF24-0920 entitled ``Novel magnets through interdisciplinarity and nanocomposites'' and the Independent Research Fund Denmark grant ``The next generation permanent magnet for the green transition'', 4307-00105B.

\bibliographystyle{IEEEtran}

\begin{thebibliography}{10}

\bibitem{Zhu_1993}
Z.~Q. Zhu, D.~Howe, E.~Bolte, and B.~Ackermann, ``Instantaneous magnetic field distribution in brushless permanent magnet {DC} motors. {I}. {O}pen-circuit field,'' {\em IEEE Transactions on Magnetics}, vol.~29, no.~1, pp.~124--135, 1993.

\bibitem{Markovic_2009}
M.~Markovic and Y.~Perriard, ``Optimization design of a segmented {H}albach permanent-magnet motor using an analytical model,'' {\em IEEE Transactions on Magnetics}, vol.~45, no.~7, pp.~2955--2960, 2009.

\bibitem{Lubin_2010}
T.~Lubin, S.~Mezani, and A.~Rezzoug, ``Analytical computation of the magnetic field distribution in a magnetic gear,'' {\em IEEE Transactions on Magnetics}, vol.~46, no.~7, pp.~2611--2621, 2010.

\bibitem{Ravaud_2009a}
R.~Ravaud, G.~Lemarquand, and V.~Lemarquand, ``Force and stiffness of passive magnetic bearings using permanent magnets. {P}art 1: Axial magnetization,'' {\em IEEE Transactions on Magnetics}, vol.~45, no.~7, pp.~2996--3002, 2009.

\bibitem{Moresi_2003}
G.~Moresi and R.~Magin, ``Miniature permanent magnet for table-top {NMR},'' {\em Concepts in Magnetic Resonance Part B: Magnetic Resonance Engineering}, vol.~19, no.~1, pp.~35--43, 2003.

\bibitem{Raich_2004}
H.~Raich and P.~Bl{\"u}mler, ``Design and construction of a dipolar {H}albach array with a homogeneous field from identical bar magnets: {NMR} {M}andhalas,'' {\em Concepts in Magnetic Resonance Part B: Magnetic Resonance Engineering}, vol.~23, no.~1, pp.~16--25, 2004.

\bibitem{Moskowitz_1966}
R.~Moskowitz and E.~Della~Torre, ``Theoretical aspects of demagnetization tensors,'' {\em IEEE Transactions on Magnetics}, vol.~2, no.~4, pp.~739--744, 1966.

\bibitem{Osborn_1945}
J.~A. Osborn, ``Demagnetizing factors of the general ellipsoid,'' {\em Physical Review}, vol.~67, no.~11-12, pp.~351--357, 1945.

\bibitem{Tejedor_1995}
M.~Tejedor, H.~Rubio, L.~Elbaile, and R.~Iglesias, ``External fields created by uniformly magnetized ellipsoids and spheroids,'' {\em IEEE Transactions on Magnetics}, vol.~31, no.~1, pp.~830--836, 1995.

\bibitem{Joseph_1965}
R.~I. Joseph and E.~Schl{\"o}mann, ``Demagnetizing field in nonellipsoidal bodies,'' {\em Journal of Applied Physics}, vol.~36, no.~5, pp.~1579--1593, 1965.

\bibitem{Smith_2010}
A.~Smith, K.~K. Nielsen, D.~V. Christensen, C.~R.~H. Bahl, R.~Bj{\o}rk, and J.~Hattel, ``The demagnetizing field of a nonuniform rectangular prism,'' {\em Journal of Applied Physics}, vol.~107, no.~10, p.~103910, 2010.

\bibitem{Kraus_1973}
L.~Kraus, ``The demagnetization tensor of a cylinder,'' {\em Czechoslovak Journal of Physics B}, vol.~23, no.~5, pp.~512--519, 1973.

\bibitem{Caciagli_2018}
A.~Caciagli, R.~J. Baars, A.~P. Philipse, and B.~W.~M. Kuipers, ``Exact expression for the magnetic field of a finite cylinder with arbitrary uniform magnetization,'' {\em Journal of Magnetism and Magnetic Materials}, vol.~456, pp.~423--432, 2018.

\bibitem{Nielsen_2020}
K.~K. Nielsen and R.~Bj{\o}rk, ``The magnetic field from a homogeneously magnetized cylindrical tile,'' {\em Journal of Magnetism and Magnetic Materials}, vol.~507, p.~166799, 2020.

\bibitem{Slanovc_2022}
F.~Slanovc, M.~Ortner, M.~Moridi, C.~Abert, and D.~Suess, ``Full analytical solution for the magnetic field of uniformly magnetized cylinder tiles,'' {\em Journal of Magnetism and Magnetic Materials}, vol.~559, p.~169482, 2022.

\bibitem{Nielsen_2026}
K.~K. Nielsen and R.~Bj{\o}rk, ``The magnetic field of a homogeneously magnetized circular piece.'' submitted, 2026.

\bibitem{Nielsen_2019}
K.~K. Nielsen, A.~R. Insinga, and R.~Bj{\o}rk, ``The stray and demagnetizing field of a homogeneously magnetized tetrahedron,'' {\em IEEE Magnetics Letters}, vol.~10, p.~2110205, 2019.

\bibitem{Beleggia_2009}
M.~Beleggia, D.~Vokoun, and M.~De~Graef, ``Demagnetization factors for cylindrical shells and related shapes,'' {\em Journal of Magnetism and Magnetic Materials}, vol.~321, no.~9, pp.~1306--1315, 2009.

\bibitem{PratCamps_2016}
J.~Prat-Camps, C.~Navau, A.~Sanchez, and D.-X. Chen, ``Demagnetizing factors for a hollow sphere,'' {\em IEEE Magnetics Letters}, vol.~7, p.~1300104, 2016.

\bibitem{Beleggia_2003}
M.~Beleggia and M.~De~Graef, ``On the computation of the demagnetization tensor field for an arbitrary particle shape using a {F}ourier space approach,'' {\em Journal of Magnetism and Magnetic Materials}, vol.~263, no.~1-2, pp.~L1--L9, 2003.

\bibitem{James_2026}
B.~James, S.~Pollok, J.~Frellsen, and R.~Bj{\o}rk, ``The magnetic scalar potential for a rectangular prism,'' {\em IEEE Transactions on Magnetics}, vol.~62, p.~7000206, 2026.

\bibitem{Bjork_2025}
R.~Bj{\o}rk, ``The magnetic scalar potential and demagnetization vector for a cylinder tile,'' {\em Journal of Magnetism and Magnetic Materials}, vol.~632, p.~173519, 2025.

\bibitem{Nguyen_2020a}
V.~T. Nguyen, ``Magnetic field distribution of a conical permanent magnet with an application in magnetic resonance imaging,'' {\em Journal of Magnetism and Magnetic Materials}, vol.~498, p.~166136, 2020.

\bibitem{Hart_2020}
S.~Hart, K.~Hart, and J.~P. Selvaggi, ``Analytical expressions for the magnetic field from axially magnetized and conically shaped permanent magnets,'' {\em IEEE Transactions on Magnetics}, vol.~56, no.~7, pp.~1--9, 2020.

\bibitem{Nguyen_2020b}
V.~T. Nguyen, ``Analytical computation of the magnetic field of a conical permanent magnet with arbitrarily uniform magnetization,'' {\em AIP Advances}, vol.~10, p.~045208, 2020.

\bibitem{Smith_2021}
M.~Smith, N.~Fokas, K.~Hart, S.~Babic, and J.~P. Selvaggi, ``The magnetic field produced from a conical current sheet and from a thin and tightly-wound conical coil,'' {\em Progress In Electromagnetics Research B}, vol.~90, pp.~1--20, 2021.

\bibitem{Batishchev_2023}
O.~Batishchev and A.~Hyde, ``Optimization of on-axis magnetic field of uniformly magnetized cones,'' {\em Journal of Applied Physics}, vol.~134, p.~043905, 2023.

\bibitem{Nguyen_2025}
V.~T. Nguyen and T.~P.~V. Nguyen, ``Modelling and analysis of magnetic field distribution generated by a magnetised paraboloid,'' {\em Progress In Electromagnetics Research C}, vol.~157, pp.~27--36, 2025.

\bibitem{Ivkovic_2025}
N.~Ivkovi{\'c}, I.~Jovanovi{\'c}, A.~Vu{\v{c}}kovi{\'c}, and M.~Peri{\'c}, ``Magnetic field calculation of a permanent ring magnet with a trapezoidal cross-section,'' in {\em Proceedings of the 60th International Scientific Conference on Information, Communication and Energy Systems and Technologies (ICEST)}, (Ohrid, North Macedonia), 2025.

\bibitem{Tandon_2004}
S.~Tandon, M.~Beleggia, Y.~Zhu, and M.~De~Graef, ``On the computation of the demagnetization tensor for uniformly magnetized particles of arbitrary shape. {Part I}: {Analytical} approach,'' {\em Journal of Magnetism and Magnetic Materials}, vol.~271, no.~1, pp.~9--26, 2004.

\bibitem{Phatak_2011}
C.~Phatak, R.~Pokharel, M.~Beleggia, and M.~De~Graef, ``On the magnetostatics of chains of magnetic nanoparticles,'' {\em Journal of Magnetism and Magnetic Materials}, vol.~323, pp.~2912--2922, 2011.

\bibitem{Prozorov_2018}
R.~Prozorov and V.~G. Kogan, ``Effective demagnetizing factors of diamagnetic samples of various shapes,'' {\em Physical Review Applied}, vol.~10, p.~014030, 2018.

\bibitem{MagTense}
R.~Bj{\o}rk and K.~K. Nielsen, ``{MagTense} -- a micromagnetism and magnetostatic framework.'' Technical University of Denmark, DTU Energy, doi:10.11581/DTU:00000071, \url{https://www.magtense.org}, 2019.

\bibitem{Bjork_2021}
R.~Bj{\o}rk, E.~B. Poulsen, K.~K. Nielsen, and A.~R. Insinga, ``{MagTense}: A micromagnetic framework using the analytical demagnetization tensor,'' {\em Journal of Magnetism and Magnetic Materials}, vol.~535, p.~168057, 2021.

\bibitem{Jackson_1999}
J.~D. Jackson, {\em Classical Electrodynamics}.
\newblock Hoboken, NJ: Wiley, 3rd~ed., 1999.

\bibitem{Abramowitz_1964}
M.~Abramowitz and I.~A. Stegun, {\em Handbook of Mathematical Functions}.
\newblock Washington: National Bureau of Standards, 1964.

\bibitem{Lewin_1981}
L.~Lewin, {\em Polylogarithms and Associated Functions}.
\newblock New York: North-Holland, 1981.

\bibitem{Sakurai_2007}
S.~Sakurai, N.~Soda, M.~Kobayashi, and G.~Rowlands, ``Geometries with the demagnetizing energy independent of the direction of magnetization,'' {\em IEEE Transactions on Magnetics}, vol.~43, no.~3, pp.~982--991, 2007.

\bibitem{Comsol}
{COMSOL AB}, ``{COMSOL Multiphysics}.'' Stockholm, Sweden, \url{www.comsol.com}, 2023.

\end{thebibliography}

\end{document}